\documentclass[preprint2]{aastex631}

\usepackage{amsmath}
\usepackage{xspace}
\usepackage{graphicx}
\usepackage{epstopdf}
\usepackage{placeins}
\usepackage{tabularx, booktabs}

\newcommand{\twco}{$^{12}$CO\xspace}
\newcommand{\thco}{$^{13}$CO\xspace}

\newcommand{\mn}{N$_2$\xspace}

\newcommand{\coqcr}{$^{12}$CO/$^{13}$CO\xspace}
\newcommand{\hcnqcr}{H$^{12}$CN/H$^{13}$CN\xspace}
\newcommand{\hcopqcr}{H$^{12}$CO$^+$/H$^{13}$CO$^+$\xspace}
\newcommand{\cchqcr}{$^{12}$C$_2$H/C$^{13}$CH\xspace}

\newcommand{\cpqcr}{$^{12}$C$^+$/$^{13}$C$^+$\xspace}
\newcommand{\ecqcr}{[$^{12}$C/$^{13}$C]$_{\rm elem}$\xspace}
\newcommand{\cele}{[C]$_{\rm elem}$}
\newcommand{\ector}{[C/O]$_{\rm elem}$}
\newcommand{\gctor}{[C/O]$_{\rm gas}$}

\shorttitle{Carbon Isotope in the PPDs}
\shortauthors{Lee et al.}
\graphicspath{{./}{figures/}}
\begin{document}

\title{Carbon isotope chemistry in protoplanetary disks: Effects of C/O ratios}

\author[0000-0002-0226-9295]{Seokho Lee}
\affiliation{Korea Astronomy and Space Science Institute 776 Daedeok-daero, Yuseong-gu, Daejeon 34055, Republic of Korea}
\email{seokholee@kasi.re.kr}
\author[0000-0002-7058-7682]{Hideko Nomura}
\affiliation{National Astronomical Observatory of Japan, 2-21-1 Osawa, Mitaka, Tokyo 181-8588, Japan}
\author[0000-0002-2026-8157]{Kenji Furuya}
\affiliation{National Astronomical Observatory of Japan, 2-21-1 Osawa, Mitaka, Tokyo 181-8588, Japan}

%\collaboration{6}{(AAS Journals Data Editors)}

%% Note that the \and command from previous versions of AASTeX is now
%% depreciated in this version as it is no longer necessary. AASTeX 
%% automatically takes care of all commas and "and"s between authors names.

%% AASTeX 6.31 has the new \collaboration and \nocollaboration commands to
%% provide the collaboration status of a group of authors. These commands 
%% can be used either before or after the list of corresponding authors. The
%% argument for \collaboration is the collaboration identifier. Authors are
%% encouraged to surround collaboration identifiers with ()s. The 
%% \nocollaboration command takes no argument and exists to indicate that
%% the nearby authors are not part of surrounding collaborations.

%% Mark off the abstract in the ``abstract'' environment. 
\begin{abstract}
Carbon isotope fractionation of CO has been reported in the disk around TW Hya, where elemental carbon is more abundant than elemental oxygen (\ector\,$>$\,1). We investigated the effects of the \ector\, ratio on carbon fractionation using astrochemical models that incorporate isotope-selective photodissociation and isotope-exchange reactions. The \coqcr\, ratio could be lower than the elemental carbon isotope ratio due to isotope exchange reactions when the \ector\, ratio exceeds unity. The observed \coqcr\ and \hcnqcr\, ratios around TW Hya could be reproduced when the \ector\,ratio is 2--5. In the vicinity of the lower boundary of the warm molecular layer, the formation of ices leads to the gas phase \ector\, ratio approaching unity, irrespective of the total (gas + ice) \ector\, ratio.  This phenomenon reduces the variation in the \coqcr\,ratio across different \ector\, ratios.
\end{abstract}

%% Keywords should appear after the \end{abstract} command. 
%% The AAS Journals now uses Unified Astronomy Thesaurus concepts:
%% https://astrothesaurus.org
%% You will be asked to selected these concepts during the submission process
%% but this old "keyword" functionality is maintained in case authors want
%% to include these concepts in their preprints.
%\keywords{}

%% From the front matter, we move on to the body of the paper.
%% Sections are demarcated by \section and \subsection, respectively.
%% Observe the use of the LaTeX \label
%% command after the \subsection to give a symbolic KEY to the
%% subsection for cross-referencing in a \ref command.
%% You can use LaTeX's \ref and \label commands to keep track of
%% cross-references to sections, equations, tables, and figures.
%% That way, if you change the order of any elements, LaTeX will
%% automatically renumber them.
%%
%% We recommend that authors also use the natbib \citep
%% and \citet commands to identify citations.  The citations are
%% tied to the reference list via symbolic KEYs. The KEY corresponds
%% to the KEY in the \bibitem in the reference list below. 

\section{Introduction} \label{sec:intro}
The isotopic ratio of molecules is a powerful tool for investigating the origin of solar system materials and for revealing the possible chemical link between the solar system and the interstellar medium (ISM) \citep[see, e.g.,][for a recent review]{Ceccarelli2014,Nomura2023}. The carbon isotope ratios are reported to be roughly constant in the Solar System \citep{Clayton2004}. The \coqcr\,ratio in the solar photosphere \citep[93.48\,$\pm$\,0.68,][]{Lyons2018} is slightly higher than terrestrial carbonates \citep[VPDB, 88.99]{Craig1957, Fleisher2021}, while \coqcr\,ratio in comet 67P/Churyumov-Gerasimenko  is 86\,$\pm$\,8 \citep{Rubin2017}.

On the other hand, the carbon isotope ratio of CO, one of the main reservoirs of carbon, evolves in star-forming regions. In the early stages, no significant carbon fractionation is observed in both the gas and ice phases of CO \citep{Boogert2002,Pontoppidan2003,Agundez2019,Yoshida2019}. 
However, the solar-mass young stellar objects exhibit \coqcr ranging from $\sim$85 to 160 when using the near-infrared CO absorption lines \citep{Smith2015}. These values notably higher than the \coqcr\, ratios of ISM \citep[62\,$\pm$\,4,][]{Langer1993}. Intriguingly, divergent ratios of 31$_{-10}^{+17}$ (TYC8998-760-1b) and 10.2--42.6 (WASP-77Ab) have been reported in the atmosphere of young super-Jupiter \citep{Zhang2021,Line2021}. 

The carbon isotope fractionation could occur through the isotope exchange reactions  \citep[e.g.,][]{Watson1976,Langer1984,Roueff2015,Colzi2020,Loison2020} and the isotope-selective photodissociation of CO \citep[e.g.,][]{Bally1982,Visser2009}. The former is efficient in the surface layer of cloud and disk irradiated by the interstellar and stellar UV radiation, while the latter works in the cold regions \citep[e.g.,][]{Rollig2013,Furuya2018,Visser2018}.

The \coqcr\, ratio in the protoplanetary disk differs between observations and previous chemical models.
The carbon isotope fractionation of CO has been spatially observed in the disk around TW Hya \citep{Zhang2017, Yoshida2022}. On the other hand, chemical models exhibit that the \coqcr\,ratio is similar to the the elemental carbon isotope (\ecqcr) ratio \citep{Wood2009,Roueff2015,Viti2020}. The chemical models used a typical ISM values like elemental carbon abundance relative to total hydrogen (\cele) of $\sim$10$^{-4}$ and elemental carbon to oxygen (\ector) ratio lower than unity. However, the observations toward disk sources show  an elevated \ector\,ratio \citep[$\ge$ 1.0, e.g.,][]{Bergin2016} and the depletion of CO is also observed \citep[e.g.,][]{Kama2016}. 

The \ector\,ratio higher than unity could change the \coqcr\,ratio \citep{Yoshida2022}.
Therefore, in this work, we will investigate the effect of \ector\,ratio as well as \cele\ on the carbon isotope ratios in the disk.  We will briefly describe our model in Section~\ref{sec:model}. In Section~\ref{sec:results}, we will present our results, and in Section~\ref{sec:discussions}, we will compare our results with the observations. Then, we will summarize our conclusions in Section~\ref{sec:summary}.

%
% Model
%
\section{Models}\label{sec:model}
We investigate carbon isotope features in the protoplanetary disk around TW Hya using the axisymmetric two-dimensional (2\,D) thermochemical disk model, packages of unified modeling for radiative transfer, gas energetics, and chemistry \citep[PURE-C,][]{Lee2021}. The code calculates the gas and dust temperatures and the chemical abundances self-consistently for given (gas and dust) density profiles and the radiation fields.  

\subsection{Physical parameters of disk}
\begin{figure*}[t]
\includegraphics[width=0.5\textwidth]{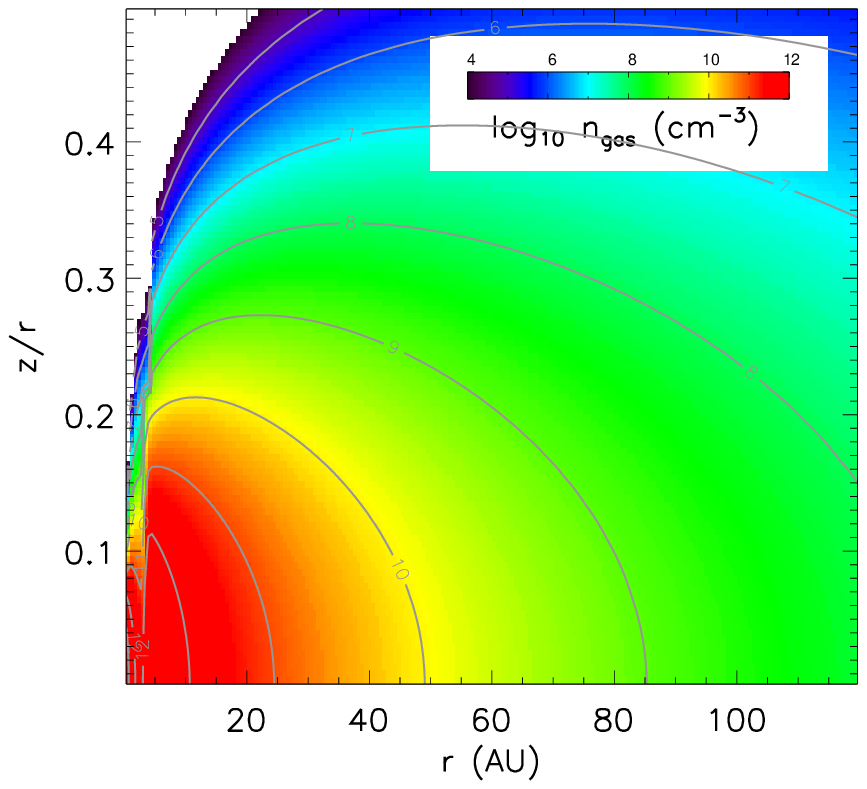}
\includegraphics[width=0.5\textwidth]{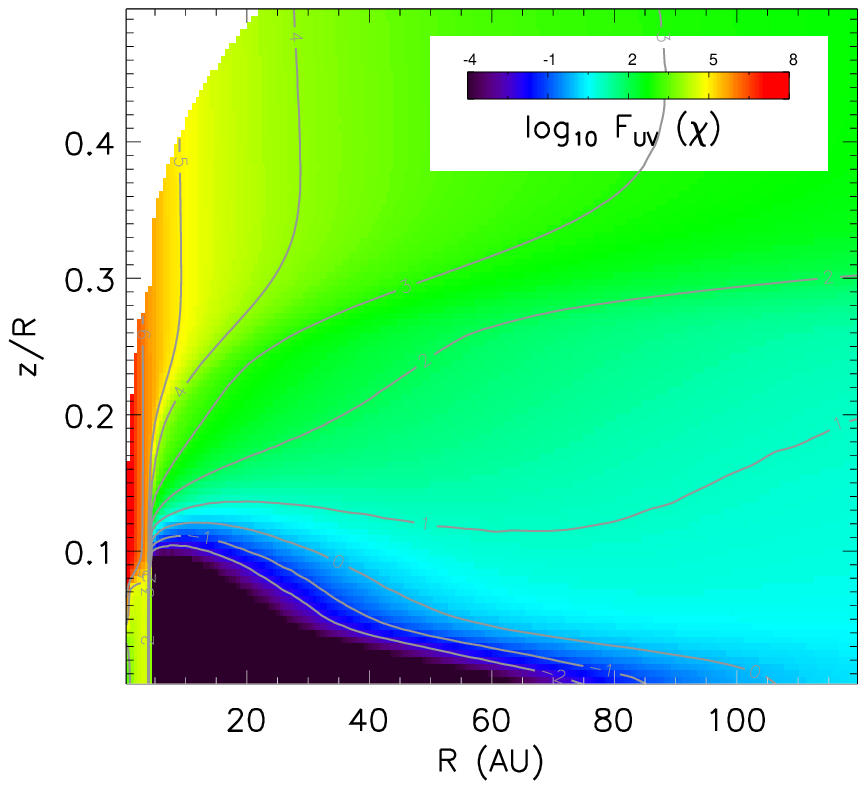}
\includegraphics[width=0.5\textwidth]{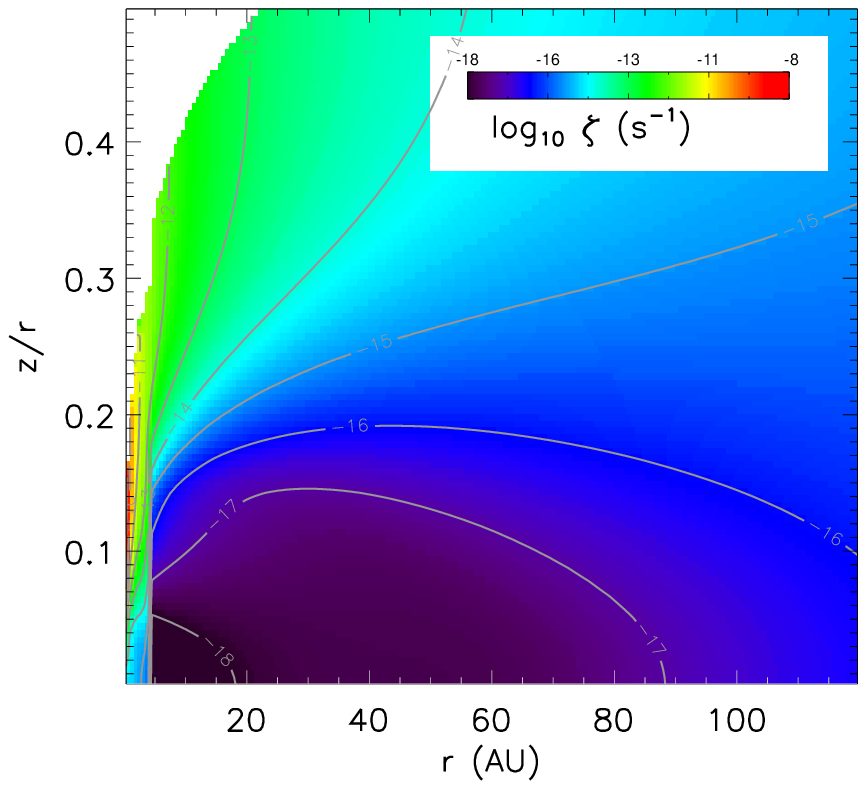}
\includegraphics[width=0.5\textwidth]{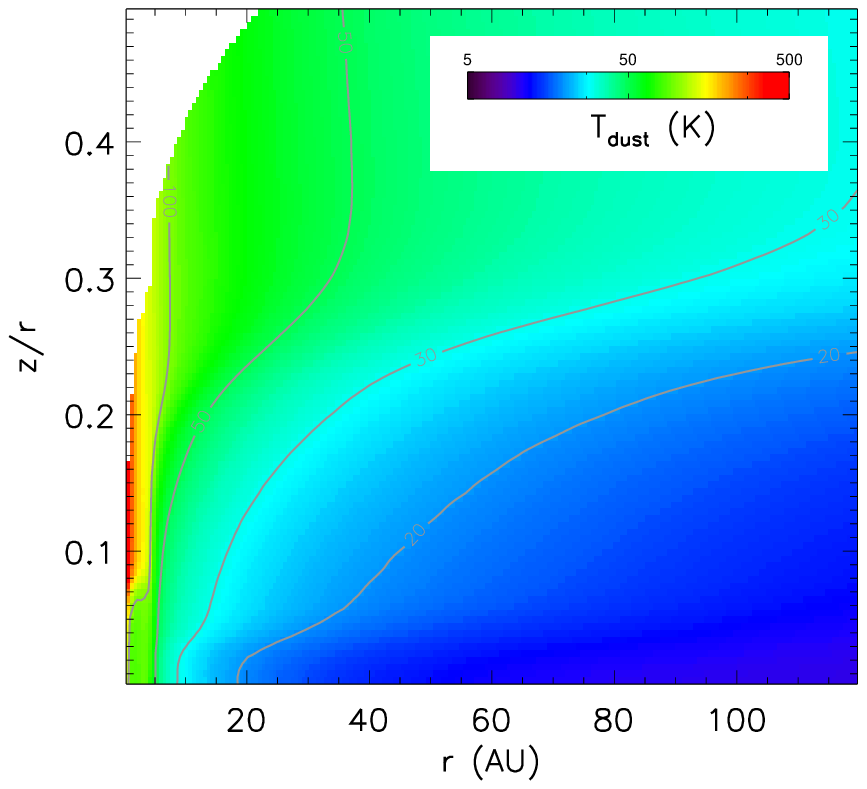}
%\vspace{-1cm}
\caption{Physical parameters. gas number density ($n_{\rm gas}$), UV flux (F$_{\rm UV}$), Ionization rates (X ray $+$ Cosmic ray,$\zeta$), and dust temperature (T$_{\rm dust}$).}
\label{fig:phys}
\end{figure*}

The density profiles of the gas and dust grains are adopted from the model of the TW Hya disk \citep{Kama2016}, which could reproduce the spectral energy distributions and the observed fluxes of multiple lines. The gas mass is set to 2.3\,$\times$\,10$^{-2} M_\odot$ and the gas to dust mass ratio is set to 200. The small dust grains (0.005\,$\mu$m--1\,$\mu$m) are coupled with gas and their mass is only 1\,\% of the total grain masses, while the large dust grains ($r_g$\,=\,0.005\,$\mu$m--1\,mm) are concentrated near the midplane and their scale height is 0.2\,$\times$ the scale height of the gas. Representative grain sizes of 0.1\,$\mu$m and 10\,$\mu$m, respectively, and density of 2.09 g\,cm$^{-3}$ are used for the chemical model \citep{Woitke2016}. The far-ultraviolet (UV) and X-ray luminosities ($L_{\rm UV}$ and $L_{\rm X}$) are 0.017\,$L_\odot$ and 1.4\,$\times$\,$10^{30}$\,erg\,s$^{-1}$, respectively. The cosmic ray ionization rate is 5\,$\times$\,10$^{-19}$\,s$^{-1}$ \citep{Kama2016}. The gas number density and the resultant UV flux normalized by Draine field \citep{Draine1978}, ionization ra/tes (X-ray and Cosmic ray), and dust temperature are shown in Figure~\ref{fig:phys}. The gas temperature decreases as the initial elemental abundances of carbon and oxygen increase, resulting in more efficient cooling, as presented in Figure~\ref{fig:all_2d_tk}.

\subsection{Chemical model}
Our chemical network is based on that in \citet{Furuya2018}, but extended to include mono-$^{13}$C species, relevant isotope exchange reactions \citep[e.g.,][]{Watson1976,Langer1984,Roueff2015,Colzi2020,Loison2020}, and isotope-selective photodissociation of CO \citep[e.g.,][]{Bally1982,Visser2009}. In our network, we do not distinguish $^{13}$C isotopomer with different $^{13}$C position for simplicity; e.g., $^{13}$CCH and C$^{13}$CH are treated as the same species.
As we are interested in only the small species such as CO, HCN, HCO$^+$, we reduce the network by choosing the species used in \citet{Lee2021} with $^{13}$C-bearing species instead of $^{15}$N-bearing species.

Important isotope exchange reactions in our models are:
\begin{equation}
^{13}\mathrm{C}^+ + \mathrm{^{12}CO}\rightleftharpoons  \mathrm{^{12}C}^+ + \mathrm{^{13}CO} + 34.7 \mathrm{K}, \label{eq:1}
\end{equation}
\begin{equation}
\mathrm{^{13}CO} + \mathrm{H^{12}CO^+} \rightleftharpoons\mathrm{^{12}CO} + \mathrm{H^{13}CO^+} + 17.4 \mathrm{K}, \label{eq:2}
\end{equation}
\begin{equation}
  \mathrm{^{13}C}^+ + \mathrm{^{12}CN} \rightleftharpoons \mathrm{^{12}C}^+ + \mathrm{^{13}CN} + 31.1  \mathrm{K}, \label{eq:3}
\end{equation}
\begin{equation}
   \mathrm{^{13}C} + \mathrm{^{12}CN} \rightleftharpoons \mathrm{^{12}C} + \mathrm{^{13}CN} + 31.1 \mathrm{K} ,\label{eq:4}
\end{equation}
\begin{equation}
   \mathrm{^{13}C} + \mathrm{H^{12}CN} \rightleftharpoons \mathrm{^{12}C} + \mathrm{H^{13}CN} + 48.0 \mathrm{K}, \label{eq:5}
\end{equation}
\begin{equation}
   \mathrm{^{13}C} + \mathrm{^{12}C_2} \rightleftharpoons \mathrm{^{12}C} + \mathrm{^{13}CC} + 26.4 \mathrm{K}, \label{eq:6}
\end{equation}
\begin{equation}
  \mathrm{^{13}C}^+ + \mathrm{^{12}C_2} \rightleftharpoons \mathrm{^{12}C}^+ + \mathrm{^{13}CC} + 26.4  \mathrm{K}, \label{eq:7}
\end{equation}
 and we adopt the reaction rates in Table 1 of \citet{Roueff2015} and in Table 2 of \citet{Loison2020}. Note that the reactions with number of 12, 13, 19, and 20 in Table 2 of \citet{Loison2020} are also included in this work. Forward reactions are exothermic reactions due to the zero point energy differences between products and reactants presented as temperatures in the reactions (\ref{eq:1})--(\ref{eq:7}), and are faster than backward ones, especially in cold conditions ($\lesssim$30\,K).  
The most important isotope exchange reaction is the reaction~(\ref{eq:1}) \citep[e.g.,][]{Langer1984}, because the  CO is a main reservoir of volatile carbon \citep{Potoppidan2014}, and many C-bearing species, such as CN and HCN, formed from C$^+$. Reaction~(\ref{eq:1}) results in a $^{13}$C-enrichment of CO and the molecules formed from CO, whereas it leads to a $^{13}$C-deficiency of the molecules formed from C$^+$.

The self-shielding of H$_2$ \citep{Draine1996}, C \citep{Kamp2000}, and N$_2$ \citep{Heays2014} as well as CO \citep{Visser2009} are considered in our model. The column densities used in the self-shielding are calculated by averaging the vertical and radial column densities weighted by the UV flux along each direction \citep[see details in][]{Lee2021}.
The isotope selective photodissociation of CO is significant only near the CO photodissociation front (near the surface layers of the molecular cloud or the disk). The self-shielding of \twco and \thco makes the \coqcr\,ratio higher than the elemental carbon isotope ratio (\ecqcr), while the carbon isotope ratio of the molecules formed from C$^+$ becomes lower than \ecqcr.

In our models, gas-ice chemistry is described by the two-phase model \citep{Hasegawa1992}.
Our model takes into account gas-phase chemistry, interactions between gas and (icy) grain surfaces, and grain surface chemistry.
Note that the all chemical reactions on the grain surface work only in the top two layers because mantle layers ($>$\,2 layers) are assumed to be inert \citep{Aikawa1999,Cuppen2017}.
In addition, the photodissociation of ices is included in this work, which is assumed to have the same rates in the gas-phase. 
In the lower boundary of the warm molecular layer, CO ice can become CO$_2$ ice on dust grains, which removes the CO in the gas phase \citep{Bergin2014,Furuya2014}. Therefore, not only the gas phase C/O ratio but also carbon isotope ratios of the gas phase species can be affected. 
In this work, we use the age of TW Hya \citep[$\sim$10 Myr, e.g., ][]{Herczeg2023} as the chemical evolution time to compare our results with observed values toward TW Hya. Furthermore, we compare the results between the evolution times of 1 Myr and 10 Myr. 

\subsection{Initial abundances}
We investigate the effects of the \ector\,ratio and the \cele\ on the carbon isotope ratios of observable species in the warm molecular layer. The initial abundances are listed in Table~\ref{tab:initabun} \citep{Cleeves2015,Lee2021}. The elemental carbon isotope (\ecqcr) ratio of 69 is adopted \citep{Wilson1999}. The elemental carbon abundances, \cele\ are 1.7\,$\times$\,10$^{-6}$, 1.7\,$\times$\,10$^{-5}$, and 1.0\,$\times$\,10$^{-4}$ in the model names beginning with A, B, C, respectively. The numbers following the model names indicate the \ector\,ratios. Note that the chemistry in the warm molecular layer tends to approach or reach an equilibrium state around the typical disk age of 1 Myr). Therefore, the results are predominantly determined by the elemental carbon and oxygen abundances rather than the initial abundances of species: whether the initial carbon and oxygen are in molecular forms or atomic forms. Thus, the initial abundances are set using only water ice, CO, and C as shown in Table~\ref{tab:initabun}.

\begin{table}[t!]
\caption{Initial abundances relative to the total hydrogen nuclei.}
\label{tab:initabun}
\centering
\begin{tabular}{cccc}
\hline\hline
Species & Abundance\tablenotemark{a} & Species & Abundance\tablenotemark{a} \\
\hline
H$_2$ & $5.000(-01)$   & He & $1.400(-01)$  \\
HCN\tablenotemark{b} & $1.000(-08)$     &  NH$_3$ ice & $9.900(-06)$ \\
N & $5.100(-06)$       &  \mn & $1.000(-06)$  \\
C$^+$\tablenotemark{b} & $1.000(-09)$   &  CH$_4$\tablenotemark{b} & $1.000(-07)$  \\
CN\tablenotemark{b} & $6.600(-08)$      & H$_3^+$ & $1.000(-08)$ \\
HCO$^+$\tablenotemark{b} & $9.000(-09)$ & C$_2$H\tablenotemark{b} & $8.000(-09)$ \\
Fe$^+$ & $1.000(-11)$   & Mg$^+$ & $1.000(-11)$ \\
Si$^+$ & $1.000(-11)$  & S$^+$ & $1.000(-11)$ \\
\hline\hline
Model   & \multicolumn{3}{c} {Species / Abundances\tablenotemark{a}} \\
\cline{2-4}
name\tablenotemark{c}    &   H$_2$O ice  & CO\tablenotemark{b}            & C\tablenotemark{b} \\
\hline
A0.5      & $2.50(-06)$  &  $1.00(-06)$ & $7.00(-07)$\\
A1.0      & $8.85(-07)$  &  $1.00(-06)$ & $7.00(-07)$ \\
A1.5      & $2.50(-07)$  &  $1.00(-06)$ & $7.00(-07)$ \\
A2.0      & $4.40(-07)$  &  $5.00(-07)$ & $1.20(-06)$ \\
A5.0      & $2.75(-07)$  &  $1.00(-07)$ & $1.60(-06)$ \\

B0.5      & $2.50(-05)$  &  $1.00(-05)$ & $7.00(-06)$\\
B1.0      & $7.15(-06)$  &  $1.00(-05)$ & $7.00(-06)$ \\
B1.5      & $1.30(-06)$  &  $1.00(-05)$ & $7.00(-06)$ \\
B2.0      & $3.50(-06)$  &  $5.00(-06)$ & $1.20(-05)$ \\
B5.0      & $2.42(-06)$  &  $1.00(-06)$ & $1.60(-05)$ \\

C0.5      & $1.00(-04)$  &  $1.00(-04)$ & $7.00(-07)$\\
C1.0      & $5.00(-05)$  &  $5.00(-05)$ & $5.00(-05)$ \\
C1.5      & $1.67(-05)$  &  $5.00(-05)$ & $6.00(-05)$ \\
C2.0      & $2.50(-05)$  &  $2.50(-05)$ & $7.50(-05)$ \\
C5.0      & $1.50(-05)$  &  $5.00(-06)$ & $9.50(-05)$ \\

\hline
\end{tabular}
\tablenotetext{a}{ x(y) means x\,$\times$\,10$^y$.}
\tablenotetext{b}{ The initial abundance of $^{13}$C bearing species
  is by a factor of \\ 69 lower than that of $^{12}$C bearing species except for C$_2$H (34.5). The latter contains two carbon atoms and doubly counted. 
  \tablenotetext{c}{ The elemental carbon  abundances are 1.7\,$\times$\,10$^{-6}$,   1.7\,$\times$\,10$^{-5}$, and \\
  1.0\,$\times$\,10$^{-4}$  for the model names beginning with A, B, and C,\\ respectively. The following numbers indicate the \ector\,ratio.}
  }
\end{table}

\section{Results}\label{sec:results}

\begin{figure*}[t]
\includegraphics[width=0.9\textwidth]{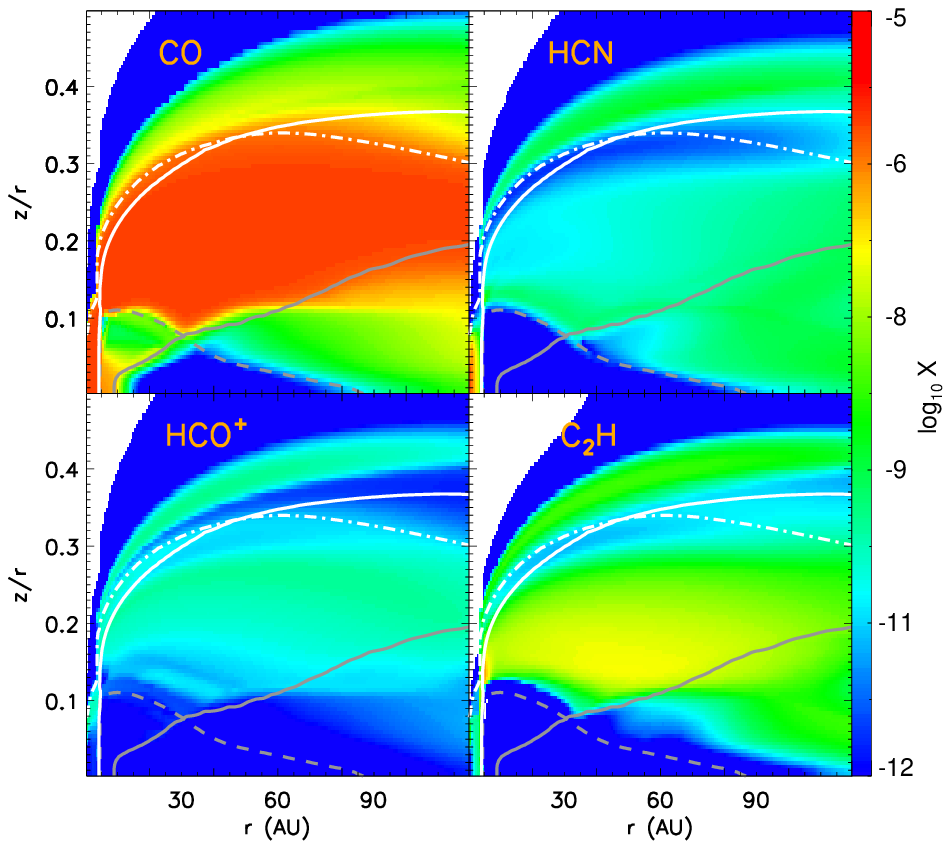}
%\vspace{-1cm}
\caption{Abundance distributions for CO, HCO$^+$, HCN, and C$_2$H in the reference model with \cele\,=\,1.7\,$\times$\,$10^{-6}$ and  \ector\,=\,2.0. The solid and dashed gray curves indicate the CO snow surface using the balancing between freeze-out onto dust grains and thermal evaporation (20--30\,K), and the dust-attenuated UV flux of 0.1 Draine Field \citep{Draine1978}. The CO photodissociation front where the CO abundance is a half of \cele, is presented in the dot-dashed curves. The white solid line indicates the gas temperature of 100\,K.}
\label{fig:abun_2d}
\end{figure*}
%\subsection{Abundances and Carbon isotope ratios}
Figure~\ref{fig:abun_2d} shows the 2D abundance distributions for CO, HCO$^+$, HCN, and C$_2$H in the reference model with 
\cele\,=\,1.7$\times$\,$10^{-6}$ and \ector\,=\,2.0, which reproduces well the observed molecular lines \citep{Kama2016}.  The gas-phase CO is abundant in the warm molecular layer \citep[e.g.,][]{Aikawa2002}, which is surrounded by the CO photodissociation front (the dot-dashed white curves where the CO abundance is half of \cele) at the upper boundary and the CO snow surface ($\sim$20--30\,K, the solid gray curves) at the lower boundary.

The lower boundary of warm molecular layer in the inner disk  ($<$\,30\,au) can be modified by the CO$_2$ ice formation. When the ionization rate is higher than $\sim$10$^{-18}$\,s$^{-1}$, the gas-phase CO is converted into CO$_2$ ice within a few Myr \citep{Bergin2014,Furuya2014,Lee2021}. Therefore, CO is depleted above the CO snow surface (the solid gray curve) as shown in the top left panel of Figure~\ref{fig:abun_2d}, where the X-ray ionization rate is higher than 10$^{-18}$\,s$^{-1}$.  In addition, CO ice reacts with OH ice, which is a product of the photodissociation of water ice, and also becomes the CO$_2$ ice  \citep{Ruaud2019,Furuya2022b}. However, in regions with the UV flux (the sum of stellar and external fields) above $\sim$0.1 Draine field, the CO$_2$ ice is photodissociated and goes back to the CO ice, which is subsequently thermally desorbed.  

Below the CO snow surface in the outer disk, 
a fraction of HCN and C$_2$H as well as CO are in the gas phase, which could contribute to the observed column density. They are photodesorbed by the UV photons when the UV flux is higher than 0.1 Draine field (the dashed curves in Figure~\ref{fig:abun_2d}), although thermal desorption is inefficient and most volatiles are in the ice phase. Note that the binding energies (on water ices) of HCN, C$_2$H, and CO$_2$ are 3700\,K, 3000\,K, and 2600\,K, respectively, which are higher than the CO binding energy of 1150\,K \citep{Wakelam2017,Furuya2018}. 

\begin{figure*}[t]
\includegraphics[width=0.9\textwidth]{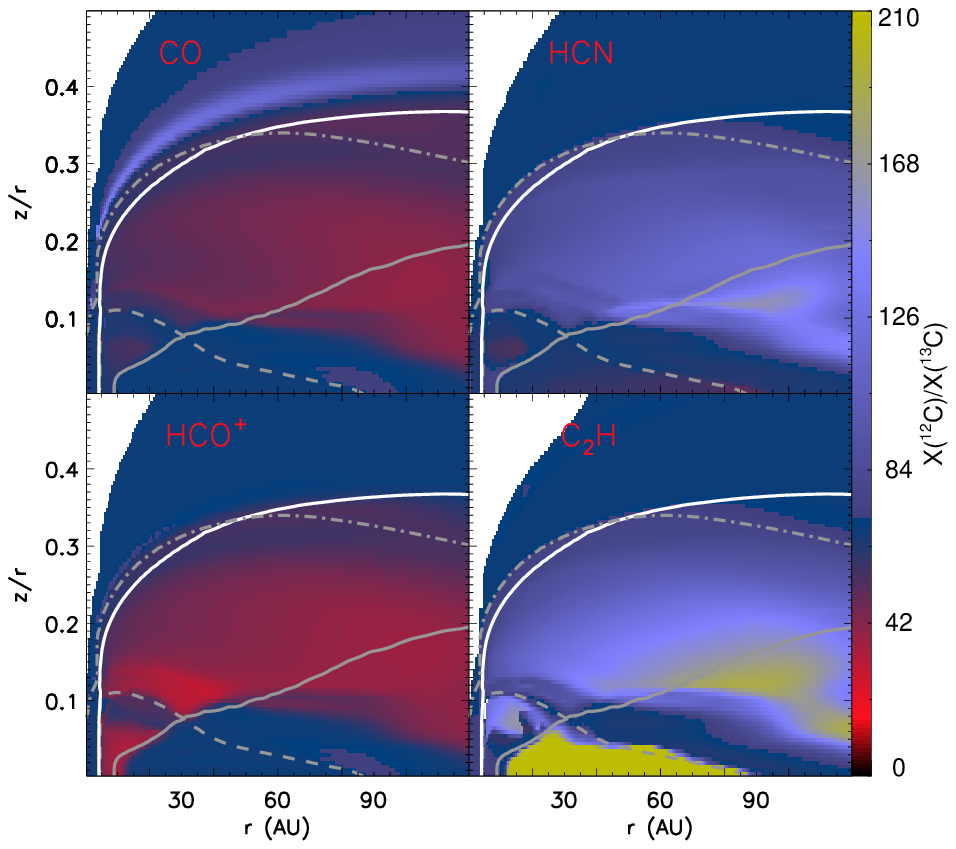}
%\vspace{-1cm}
\caption{Carbon isotope distributions for the same molecules in Figure~\ref{fig:abun_2d}. The curves correspond to those in Figure~\ref{fig:abun_2d}. Note that the carbon isotope ratio of C$_2$H is derived as 2\,$\times$\, n(C$_2$H)/n(C$^{13}$CH).}
\label{fig:iso_2d}
\end{figure*}

%[Rewrite 2 paragraphs.]
Figure~\ref{fig:iso_2d} shows the 2D distributions of the carbon isotope ratios for the same molecules depicted in Figure~\ref{fig:abun_2d}, and the vertical distributions of the abundances and the isotope ratios at $r=$\,30\,au are displayed in the right panels in Figure~\ref{fig:vertical_gas}. 
In the UV-dominated surface layers of the disk, the isotope selective photodissociation of CO determines \coqcr\,ratio \citep[e.g.,][]{Visser2018}.  All CO molecules are efficiently photodissociated, resulting in a very small abundance of CO, thus, the ratio is close to \ecqcr\, near the atmosphere surface ($z/r$\,$\gtrsim$\,0.4 around $r$\,=\,30\,au). However, this ratio increases as the CO molecules begin to survive from the UV photons (light-blue color in the top left panel) in the vicinity of the CO photodissociation front \citep[e.g.,][]{Visser2009}.  Nevertheless, the isotope exchange reaction (\ref{eq:1}) partly mitigates the fractionation caused by isotope selective photodissociation of CO when the gas temperature falls below $\sim$100\,K (indicated by the solid white curves in Figure~\ref{fig:iso_2d}) near the CO photodissociation front \citep{Wood2009}. Therefore, even around the CO photodissociation front, CO is enriched in $^{13}$C as shown in the top left panel of Figure~\ref{fig:iso_2d} (see also Figure~\ref{fig:all_2d_iso_co}).

In the warm molecular layer, CO becomes enriched in $^{13}$C (red color) through the isotope exchange reaction (\ref{eq:1}). The reaction (\ref{eq:2}) induces greater carbon fractionation in HCO$^+$ than in CO. On the other hand, the molecules formed from C$^+$ (HCN and C$_2$H) are $^{13}$C-poor in the warm molecular layer as shown in the right column of Figure~\ref{fig:iso_2d}. The degree of carbon isotope fractionation of the molecules is more significant below the CO snow surface (solid gray lines) in the outer disk because the reaction (\ref{eq:1}) is more efficient in the colder region. 

The carbon isotope ratios in the disk could change with the age of the disk. As shown in the middle row of Figure~\ref{fig:vertical_gas}, 
the molecules in the lower height of the warm molecular layer mainly contribute to the (vertically integrated) observed column density. The chemistry in the warm molecular layer generally becomes an equilibrium condition within 1\,Myr. However, the CO abundance starts to decrease due to CO$_2$ ice formation after $\sim$1\,Myr \citep{Furuya2014} in the lower boundary of the warm molecular layer in the inner disk ($<$\,40\,au), where the X-ray ionization rate is as low as 10$^{-18}$--10$^{-17}$\,s$^{-1}$. Therefore, the observed carbon isotope ratio as well as the abundances could be evolved between $\sim$1--10\,Myr.

\subsection{Effects of \ector}

\begin{figure*}[t]
\includegraphics[width=0.9\textwidth]{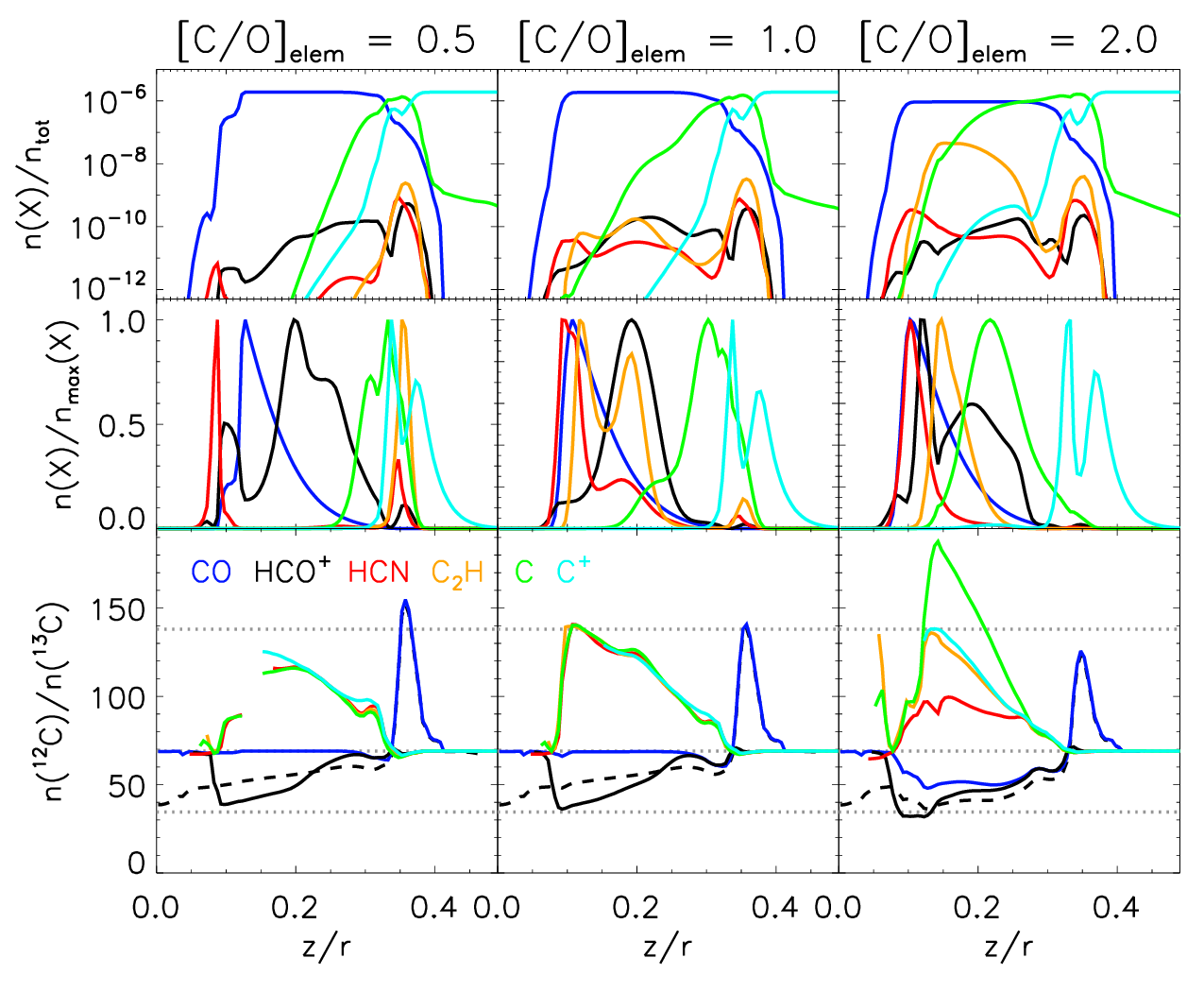}
%\vspace{-1cm}
\caption{Vertical abundance (top) and isotope ratio (bottom) distributions along the height at 30\,au in the \ector\, ratio of 0.5 (left), 1.0 (middle), and 2.0 (right) with the \cele\,=\,1.7\,$\times$\,10$^{-6}$. The relative contributions of each height to the vertically integrated column density are presented in the middle layer. The blue, black, red, orange, green, and cyan lines indicates the values for the CO, HCO$^+$, HCN, C$_2$H, C, and C$^+$, respectively. The black dashed lines in the bottom panels indicate the \hcopqcr\,ratio using Equation~\ref{eq:wood}.  The horizontal dotted grey lines present the 0.5, 1, and 2 times \ecqcr\, ratios. Note that the carbon isotope ratio of C$_2$H is derived as 2\,$\times$\,n(C$_2$H)/n(C$^{13}$CH).  }
\label{fig:vertical_gas}
\end{figure*}

\begin{figure*}[t]
\includegraphics[width=0.8\textwidth]{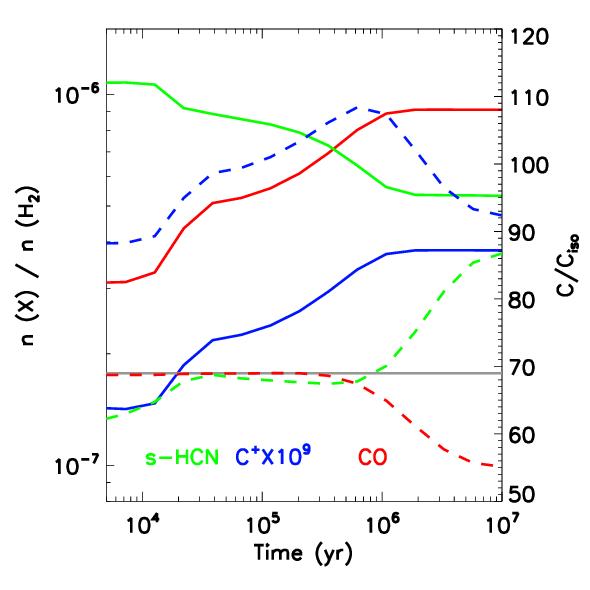}
%\vspace{-1cm}
\caption{Chemical evolution at 30\,au and $z/r$=\,0.1 in the reference model. The solid and dashed lines represent the abundances (on the left axis) and carbon isotope ratios (on the right axis), respectively. The values for CO, C$^+$, and HCN ice are plotted in the red, blue, and green lines, respectively. The C$^+$ abundance is depicted as increased by a factor of by 10$^9$.}
\label{fig:time_evol}
\end{figure*}

Figure~\ref{fig:vertical_gas} shows vertical distributions of abundance (top) and carbon isotope ratio (bottom) at the distance of 30 au from the central star when \ector\,ratios are 0.5, 1.0, and 2.0 from left to right. 
When \ector\,$\leq$\,1, most elemental carbon exists in the form of CO in the warm molecular layer (0.1\,$\le$\,z/r\,$\le$\,0.35). On the other hand, when \ector\,$>$\,1,  atomic carbon, carbon chain molecules, and cyanide (e.g., C$_2$, C$_2$H, HCN) are also abundant due to the presence of a sufficient amount of leftover elemental carbon remaining after the formation of CO  \citep[e.g.,][]{Bergin2014,Cleeves2018,Lee2021}.

The \coqcr\,ratio in the warm molecular layer depends on the \ector\,ratio, and is primarily determined by the isotope exchange reaction (\ref{eq:1}).  When the \ector\,ratio is $\le$\,1, the self-shielding is very efficient, making CO the primary reservoir for elemental carbon. Therefore, the isotope exchange reactions cannot  induce the CO fractionation, and thus, the  \coqcr\,ratio is close to \ecqcr\  \citep[see the blue lines in the left and middle columns of Figure~\ref{fig:vertical_gas},][]{Wood2009}. 
However, when the \ector\,ratio exceeds unity (right column of Figure~\ref{fig:vertical_gas}), the abundances of atomic carbon gas and ices of HCN and carbon chain molecules, such as C$_m$ and C$_m$H$_n$, become comparable to that of CO gas. Consequently, sufficient amount of C$^+$ continues to be supplied by carbon-containing species, and the isotope exchange reaction (\ref{eq:1}) results in the \coqcr\,ratio lower than \ecqcr\,ratio, as presented by the blue line in the bottom right panel of Figure~\ref{fig:vertical_gas}. 

Figure~\ref{fig:time_evol} exhibits the chemical evolution at the vicinity of the lower boundary of the warm molecular layer ($z/r\,\sim$0.1) at 30\,au in the reference model. Ices of HCN and C$_m$H$_n$ are abundant and comparable to the CO gas, and they exhibit $^{13}$C-deficiency due to $^{13}$C-deficient C$^+$, produced by the isotope exchange reaction (\ref{eq:1}), when \ector\,$>$\,1. The ices of HCN and C$_m$H$_n$ undergo a transformation to the CO gas after a few 10$^3$--10$^4$\,yrs, which is initiated by the photo-desorption of ices by stellar UV photons. The newly supplied elemental carbons (as well as elemental oxygens) in the gas phase increase the CO abundance and decrease the \coqcr\,ratio through the reaction (\ref{eq:1}).

%
% HCO+
%
Carbon isotope ratios of other molecules are also altered by the \ector\ ratio. \citet{Wood2009} showed that based on the dominance of reaction (\ref{eq:2})  \citep{Langer1984}, one can estimate the expected isotope ratio of \hcopqcr\, through the following relation:
\begin{equation} \label{eq:wood}
    \frac{n({\rm H^{12}CO^+})}{n({\rm H^{13}CO^+})} \sim \exp{\left(-\frac{9\, {\rm K}}{T_{\rm gas}}\right)}\frac{n({\rm ^{12}CO})}{n({\rm ^{13}CO})}.
\end{equation}
The dashed black lines in Figures~\ref{fig:vertical_gas} and \ref{fig:vertical_gas_el} show the \hcopqcr\,ratio using the above equation along with the \coqcr\,ratio (the solid blue lines) while the solid black lines exhibit the \hcopqcr\,ratio in our model. 
The strong gradients in physical and chemical conditions including both gas temperature, CO abundance, and electron abundance cause our models to deviate from this simple relationship \citep[see Figure~3 in ][]{Furuya2022}, and thus models will be necessary for interpreting observed \hcopqcr\ ratios in disks.

HCN and C$_2$H are formed from C$^+$, and thus, their carbon isotope ratios align with \cpqcr. However, when the \ector\,ratio exceeds unity, the atomic carbon and C$_2$ are abundant in the warm molecular layer. Thus, isotope exchange reactions \ref{eq:3}--\ref{eq:7} \citep{Loison2020} contribute to lower \hcnqcr and \cchqcr ratios compared to the \cpqcr\,ratio as shown in the bottom panel of Figure~\ref{fig:vertical_gas}.

\subsection{Effects of \cele}
\begin{figure*}[t]
\includegraphics[width=0.9\textwidth]{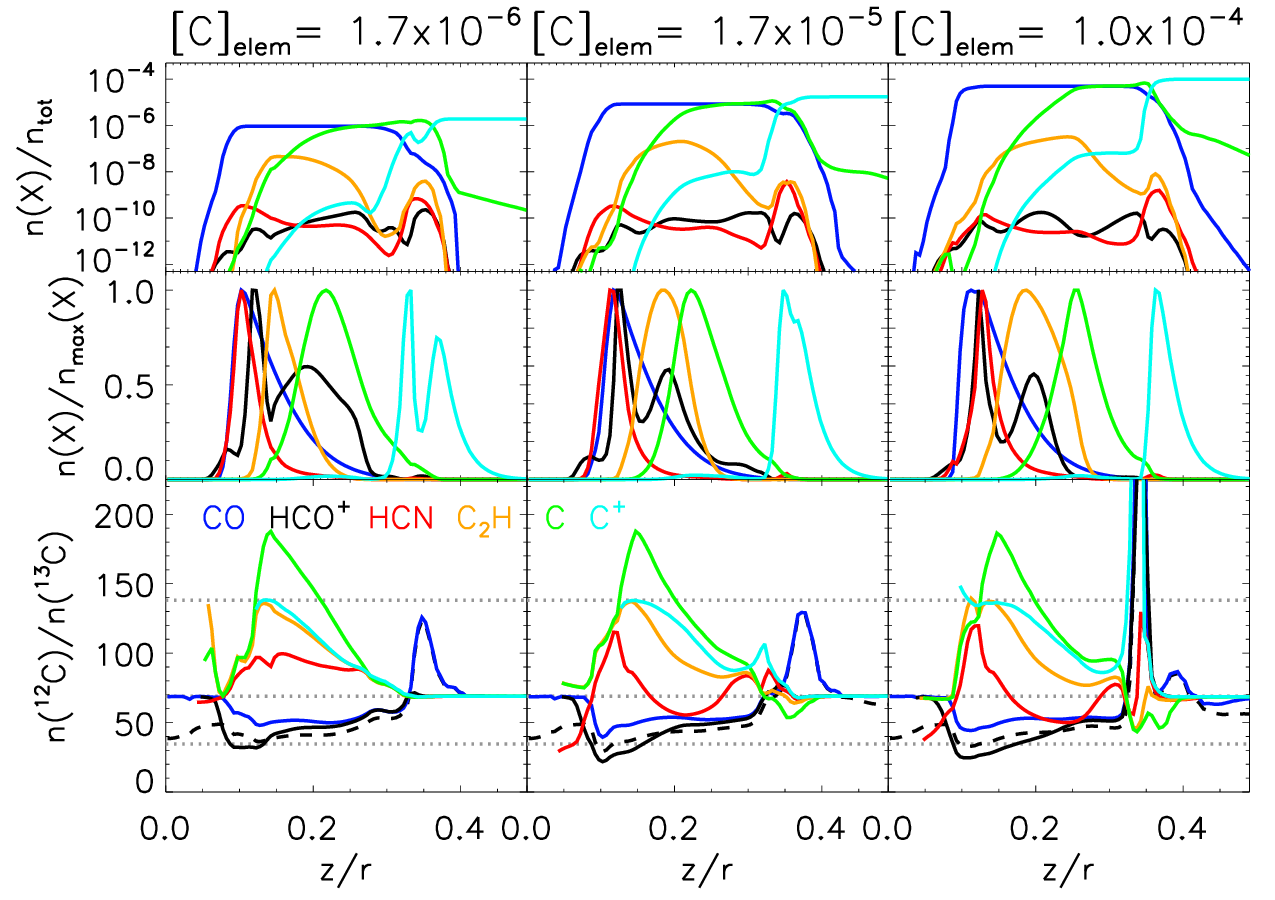}
%\vspace{-1cm}
\caption{Same as Figure~\ref{fig:vertical_gas} except for the  \cele\,=\,1.7\,$\times$\,10$^{-6}$ (left), 1.7\,$\times$\,10$^{-5}$ (middle), and 1.0\,$\times$\,10$^{-4}$ (right) with the \ector\,ratio of 2.0.}
\label{fig:vertical_gas_el}
\end{figure*}

The elemental carbon abundance (\cele) of disks is poorly constrained, and like oxygen,  it is thought to vary \citep[e.g.,][]{Ansdell2016}. In this section we explore differences in the elemental carbon abundances on the carbon isotope ratios. 
Figure~\ref{fig:vertical_gas_el} shows the differences in abundances and carbon isotope ratios for the molecules presented in Figure~\ref{fig:vertical_gas}, considering variations in  elemental carbon abundances (\cele), fixing the \ector\,ratio of 2.0. The peak abundances of CO/C/C$^+$ increase by a factor of 10 and $\sim$50 as the \cele\, increases. The height of the CO photodissociation front where the CO abundance is a half of \cele, increases with a higher \cele. This is because the CO self-shielding is more effective due to the higher CO abundance in the atmosphere. However, it is worth noting that the contribution of this height to the column density, when integrated along the entire height, is negligible as shown in the middle panels of Figure \ref{fig:vertical_gas_el}.

% Carbon Isotope ratio
Carbon isotope ratios of CO, atomic carbon, and C$^+$ in the warm molecular layer show a similar trend regardless of \cele\ as shown in the bottom panels of Figure~\ref{fig:vertical_gas_el}. In the warm molecular layer, the ratios are mainly determined by the isotope exchange reaction (\ref{eq:1}). As the \cele\ increases, the gas temperature could be decreased due to efficient cooling by CO and H$_2$O in the upper warm molecular layer ($z/r>$\,0.25; see Figure~\ref{fig:all_2d_tk}). However, the gas temperature is close to the dust temperature in the lower warm molecular layer. Therefore, CO, atomic C, and C$^+$ show similar carbon isotope ratios across different \cele\ in the warm molecular layer. 

However, in the atmosphere, the isotope selective photodissociation of CO is important, which is more efficient as  \cele\ increases.
Furthermore, gas temperature decreases according to \cele. Therefore, carbon isotope ratios exhibit complicated trends due to the competition between the isotope exchange reaction (\ref{eq:1}) and the isotope selective photodissociation of CO.   

The carbon isotope ratios of the products are determined by the contribution of the inheritance from CO or C$^+$ and the additional isotope reactions:  reaction (\ref{eq:2}) for HCO$^+$, reactions (\ref{eq:5}) and (\ref{eq:6}) for the C$_2$H, and reactions (\ref{eq:3}) and (\ref{eq:4}) for the HCN, respectively. In the warm molecular layer, the abundances of CO, C, and C$^+$ are more sensitive to the elemental carbon  abundance compared to those of HCO$^+$, HCN, and C$_2$H. Thus, the contribution of the latter isotope reactions increases, and the HCO$^+$, C$_2$H, and HCN are $^{13}$C-enriched compared to their mother species (CO and C$^+$).

\section{Discussions}\label{sec:discussions}
\subsection{Comparison with observations}
Previous disk models used the elemental abundances similar to the ISM values with the \ector\,ratio lower than unity  \citep{Wood2009,Furuya2014,Visser2018}.  In this case, the \coqcr\,ratio is similar to the \ecqcr\,ratio \citep{Roueff2015,Viti2020} or close to the initial \coqcr\,ratio \citep{Wood2009}. On the other hand, molecules formed from C$^+$ in the cold region could undergo fractionation through the reaction (\ref{eq:1}). 
However, the elemental abundances of the disk around TW Hya differ from the ISM condition. The bright C$_2$H emission is observed toward TW Hya, implying an elevated \ector\,ratio \citep[$\ge$ 1.0, e.g.,][]{Bergin2016}, and the depletion of CO is also observed \citep{Kama2016}. 
Furthermore, those features are common in the disk sources \citep{Bergin2016,Miotello2019,Bergner2020,Bosman2021a,Bosman2021b,Sturm2022}. Therefore, the observed carbon isotope ratios could be explained by elevated \ector\, ratios.

\subsubsection{CO}\label{subsec:co}
\begin{figure*}[t]
\includegraphics[width=1.0\textwidth]{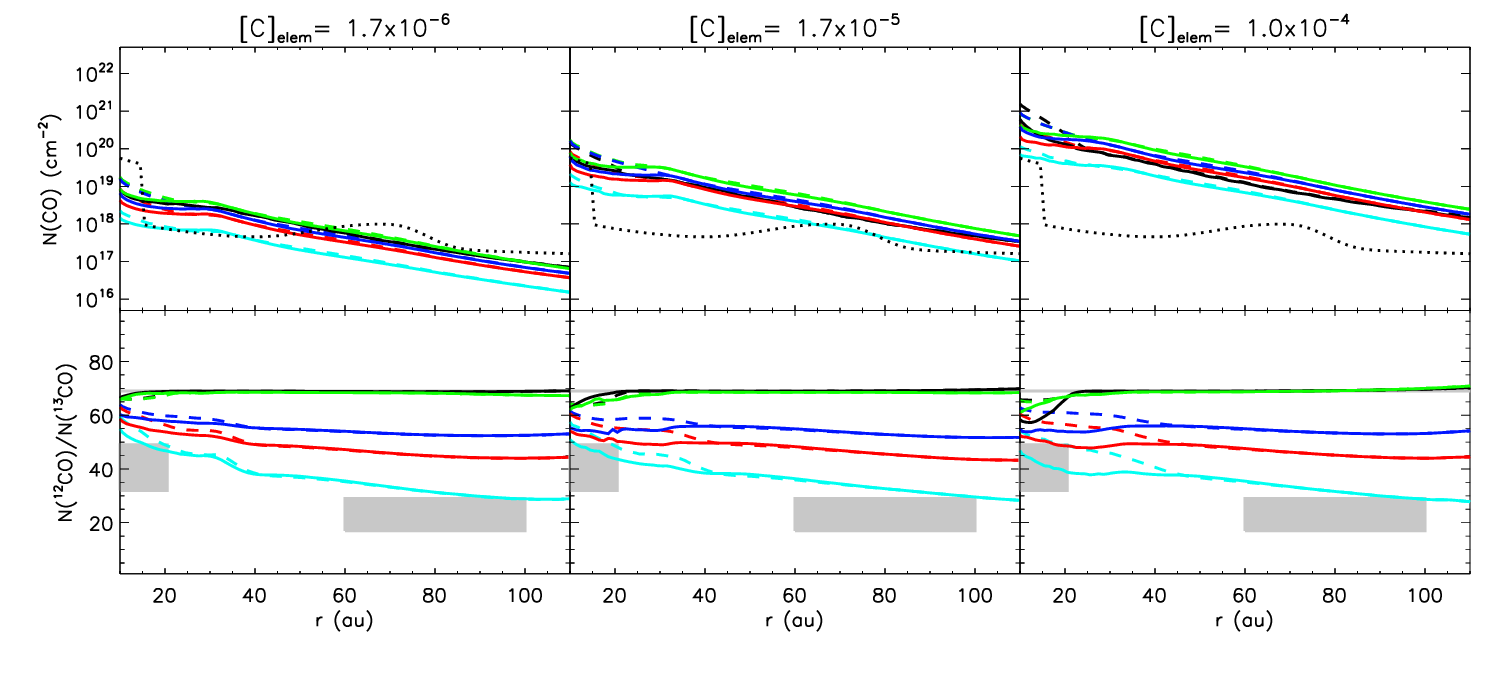}
\caption{CO column densities (top) and column density ratio of \coqcr\ along the distance from the central star. The models with the elemental carbon abundances (\cele) of 1.7\,$\times$\,10$^{-6}$, 1.7\,$\times$\,10$^{-5}$, and 1.0\,$\times$\,10$^{-4}$ are plotted in the left, middle, and right columns, respectively. The \ector\, ratios of 0.5, 1.0, 1.5, 2.0, and 5.0 are presented in the black, green, blue, red, and cyan lines. The solid and dashed lines indicate the models with the evolution times of 10 and 1 Myr, respectively. The black dotted lines in the top panels indicate the observed CO column density \citep{Huang2018}. The gray boxes in the bottom panels indicate the observed \coqcr ratios \citep{Zhang2017,Yoshida2022}. }
\label{fig:all_co}
\end{figure*}

Spatially resolved observations of the carbon isotope fractionation have been conducted in the disk around TW Hya. The observationally derived $^{12}$C$^{18}$O/$^{13}$C$^{18}$O ratio is 40$_{-6}^{+9}$  around 5--20.5 au \citep{Zhang2017} and the \coqcr\,ratio is 23$\pm$6 around 60 -- 100 au and $\sim$100 in the outer disk \citep[$>$\,100\,au, ][]{Yoshida2022}.

Figure~\ref{fig:all_co} shows the CO column densities (top panels) and the column density ratios of \twco to \thco (bottom panels) as functions of a distance from the central star. The elemental carbon  abundances, \cele\, are 1.7\,$\times$\,10$^{-6}$, 1.7\,$\times$\,10$^{-5}$, and 1.0\,$\times$\,10$^{-4}$, respectively, from left to right, respectively.
The model with \ector\,$\sim$5 reasonably well reproduces the observed \coqcr\,ratio. However, the observed \coqcr\,ratio is still slightly lower than the \coqcr\,ratio predicted by our model with \ector\,= 5 (see the solid lines in the left bottom panel of Figure~\ref{fig:all_co}), which implies that the \ector\, ratio might be even higher than five and/or the gas temperature might be lower in the TW Hya disk than that in our model. When we reran the reference model after artificially reducing the gas temperature by half, the carbon isotope ratio decreases by $\sim$15--20\%.

The \ector\ and \cele\ could affect the CO column density and carbon isotope ratio.
When \ector\,$>$\,1, the CO column density decreases due to an oxygen deficiency relative to carbon for the formation of CO. On the other hand, when \ector\,$<$\,1, the column density of CO decreases due to its conversion into CO$_2$ ice near the CO snow surface. 
The CO column densities increase as \cele\ increases, and the trends of CO column density with respect to the \ector\,ratio are similar in all three cases with varying \cele. The observed CO column density in the TW Hya disk \citep{Huang2018} appears to be similar to the models with \cele\,=\,1.7\,$\times$\,10$^{-6}$. The dotted and solid lines indicate the models with the evolution time of 1 Myr and 10 Myr, respectively. The column densities within $r=30$ au slightly decrease because the CO$_2$ ice still forms after 1 Myr. 

When \ector\,$\leq$\,1, the column density ratio of \coqcr\ is close to the \ecqcr\, ratio. However, as the \ector\,ratio exceeds unity, the column density ratio decreases below the \ecqcr\,ratio. The trends of \coqcr\ with respect to the \ector\,ratio are consistent across all three \cele\ cases. Furthermore, the \coqcr\,ratio is unaffected by the \cele. 

Our results are consistent with a simple approach in \citet{Yoshida2022}.
In the normal dense ISM condition, where elemental carbon to oxygen ratio (\ector) is $\sim$0.5 and most volatile carbon is locked up in CO, the degree of carbon fractionation in CO by reaction (\ref{eq:1}) is not significant \citep[deviation of $^{12}$CO/$^{13}$CO from \ecqcr is $\sim$30\,\% at most,][]{Furuya2011}. 
However, this situation can change when \ector\, is $\gtrsim$ 1 because, in that case, volatile carbon carriers as abundant as CO exist.
According to \citet{Yoshida2022}, the \coqcr ratio normalized by the elemental carbon isotope ratio \ecqcr\ is roughly expressed by  
\begin{eqnarray}
\label{eq:yosida}
    \mathrm{\frac{^{12}CO/^{13}CO} {[^{12}C/^{13}C]_{elem}}} = &  \\ \nonumber
    \mathrm{\frac{1}{[C/O]_{elem}}} &\mathrm{\left[\exp\left(-\frac{35 K}{T}\right)([C/O]_{elem} -1) + 1\right],}
\end{eqnarray}
assuming that the reaction (\ref{eq:1}) is in chemical equilibrium, only C$^+$ and CO are the carbon carriers, and \ector\, is larger than unity.
The above equation would give the lower limit of the normalized ratio, because other carbon carriers, such as C atoms and icy species, besides C$^+$ and CO, should exist in reality.
When \ector\, is $\lesssim \exp(35~{\rm K}/T)$ ($\sim$33 for 10\,K and $\sim$3 for 30\,K), the ratio inversely scales with \ector\,.
Therefore, $^{12}$CO/$^{13}$CO can be significantly different from \ecqcr when \ector\,$>$\,1.
According to Equation \ref{eq:yosida}, the \ector\,ratio of 3--10 is needed to reproduce the observed \coqcr ratio in the inner disk around TW Hya \citep{Yoshida2022}.

\subsubsection{HCO$^+$} \label{subsec:hcop}
\begin{figure*}[t]
\includegraphics[width=1.0\textwidth]{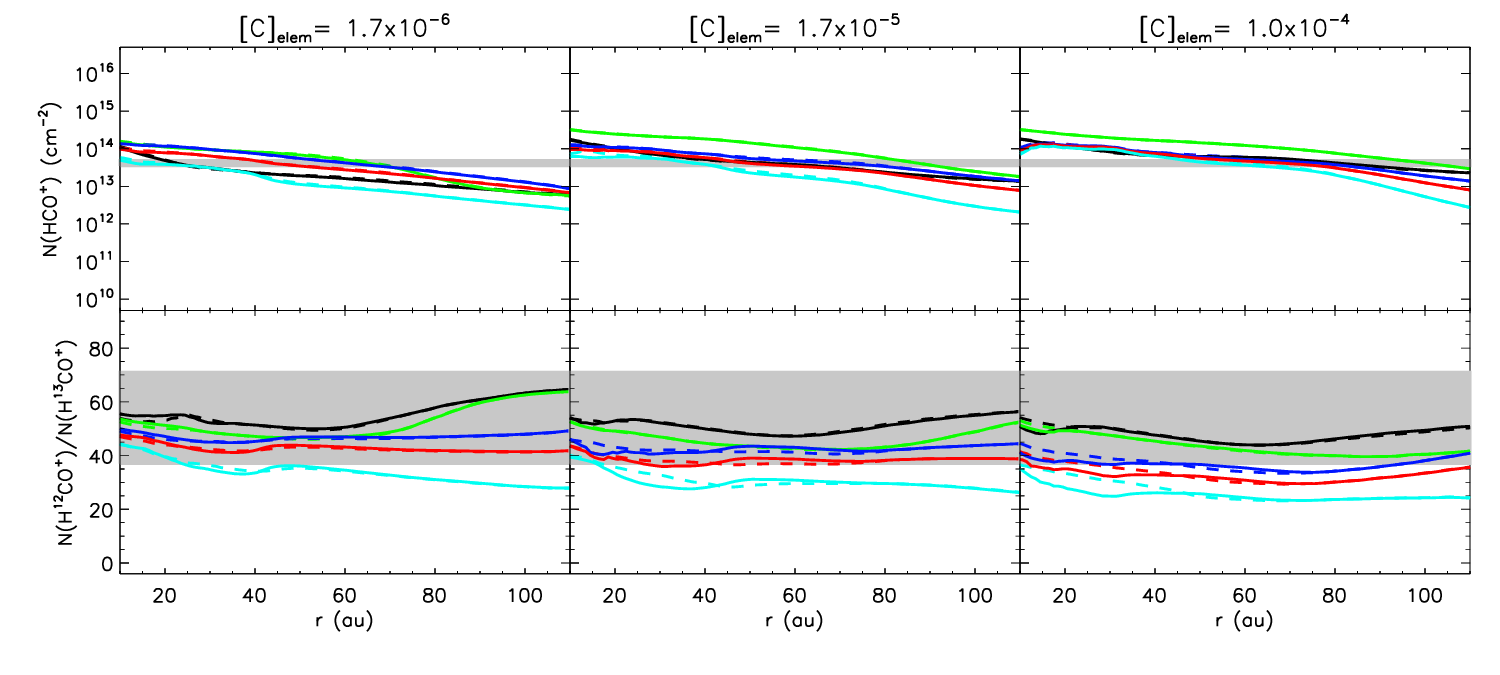}
\caption{Same as Figure~\ref{fig:all_co} except for HCO$^+$. The gray horizontal bars indicate the observed HCO$^+$ column density and \hcopqcr\,ratio \citep{Furuya2022}.
}
\label{fig:all_hcop}
\end{figure*}

The observed H$^{13}$CO$^+$ and HC$^{18}$O$^{+}$ column densities in the whole disk around TW Hya are (7.8\,$\pm$\,2.0)$\,\times$\,10$^{11}$\,cm$^{-2}$  and (7.5\,$\pm$\,1.3)\,$\times$\,10$^{10}$\,cm$^{-2}$, respectively, based on the H$^{13}$CO$^+$ and HC$^{18}$O$^+$ 4--3 lines \citep{Furuya2022}. When we assume that the H$^{12}$CO$^+$/HC$^{18}$O$^{+}$ ratio is the elemental ratio of 557 \citep{Wilson1999}, the  H$^{12}$CO$^+$ column density and the \hcopqcr\,ratio are (4.2\,$\pm$\,0.7)\,$\times$\,10$^{13}$\,cm$^{-2}$ and 54\,$\pm$\,17, respectively (see the horizontal gray boxes in Figure~\ref{fig:all_hcop}). The \ector\, ratio and \cele\, do not affect significantly the HCO$^+$ column density and the \hcopqcr\,ratio. The \hcopqcr\,ratio is lower than the \coqcr\,ratio through the isotope exchange reaction (\ref{eq:2}). This ratio is by $\sim$20\% more fractionated compared to \coqcr\, ratio in the inner disk ($<$40\,au) regardless of \ector. Most our models exhibit narrow ranges of column density and carbon isotope ratio compared to the observed ones as shown in Figure~\ref{fig:all_hcop}.

\subsubsection{HCN}\label{subsec:hcn}
\begin{figure*}[t]
\includegraphics[width=1.0\textwidth]{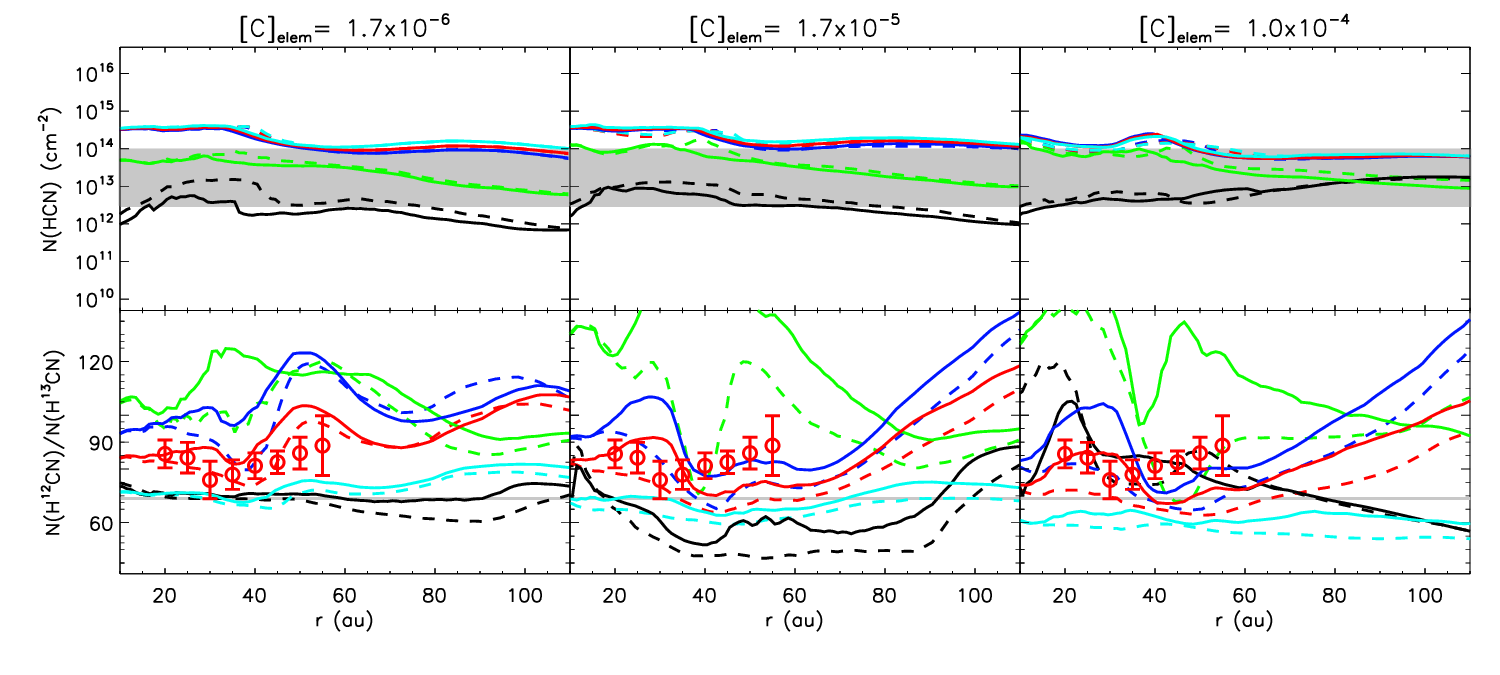}
\caption{Same as Figure~\ref{fig:all_co} except for HCN. The gray bar and red symbols are the observed column density \citep{Kastner2014} and  \hcnqcr ratios \citep{Hily-Blant2018}.}
\label{fig:all_hcn}
\end{figure*}

The \hcnqcr\,ratio is reported as $\sim$86~$\,\pm$\,4 toward TW Hya using the optically thin HCN 4--3 hyperfine lines and H$^{13}$CN 4--3 line  \citep{Hily-Blant2019}. The observed \hcnqcr\,ratio in the TW Hya disk (open circles in the bottom panels of Figure~\ref{fig:all_hcn})  is well fit with the models with \ector\,=\,2--5 (the red and cyan lines) when \cele\, is 1.7\,$\times$\,10$^{-6}$. 

Figure~\ref{fig:all_hcn} shows the HCN column densities (top panels) and the column density ratios of H$^{12}$CN to H$^{13}$CN (bottom panels). The HCN column density increases with the \ector\,ratio and is saturated when  \ector\,$\geq$\,1.5. The \hcnqcr\,ratio within $r$\,=\,80\,au increases with the \ector\,ratio and reaches its maximum value of $\geq$\,120 around the \ector\,ratio of unity. Then, the ratio decreases with increasing the \ector\,ratio when \ector\,$>$\,1. For a given \ector\,ratio, the \hcnqcr\,ratio decreases as  \cele\ increases via the isotope exchange reaction (\ref{eq:5}) as mentioned in Section~\ref{sec:results}. 
When \cele\ is higher, the lower \ector\,ratio could fit the observations. In addition, the ratio could be lower than \ecqcr\,ratio when \ector\,=\,5. The \hcnqcr\,ratio is higher at 10 Myr (solid lines) compared to 1 Myr (dashed lines), which is the opposite trend shown in the \coqcr\,ratio (see Section~\ref{subsec:co}). However, their differences are too small to distinguish observationally.

\subsubsection{C$_2$H}\label{subsec:c2h}
\begin{figure*}[t]
\includegraphics[width=1.0\textwidth]{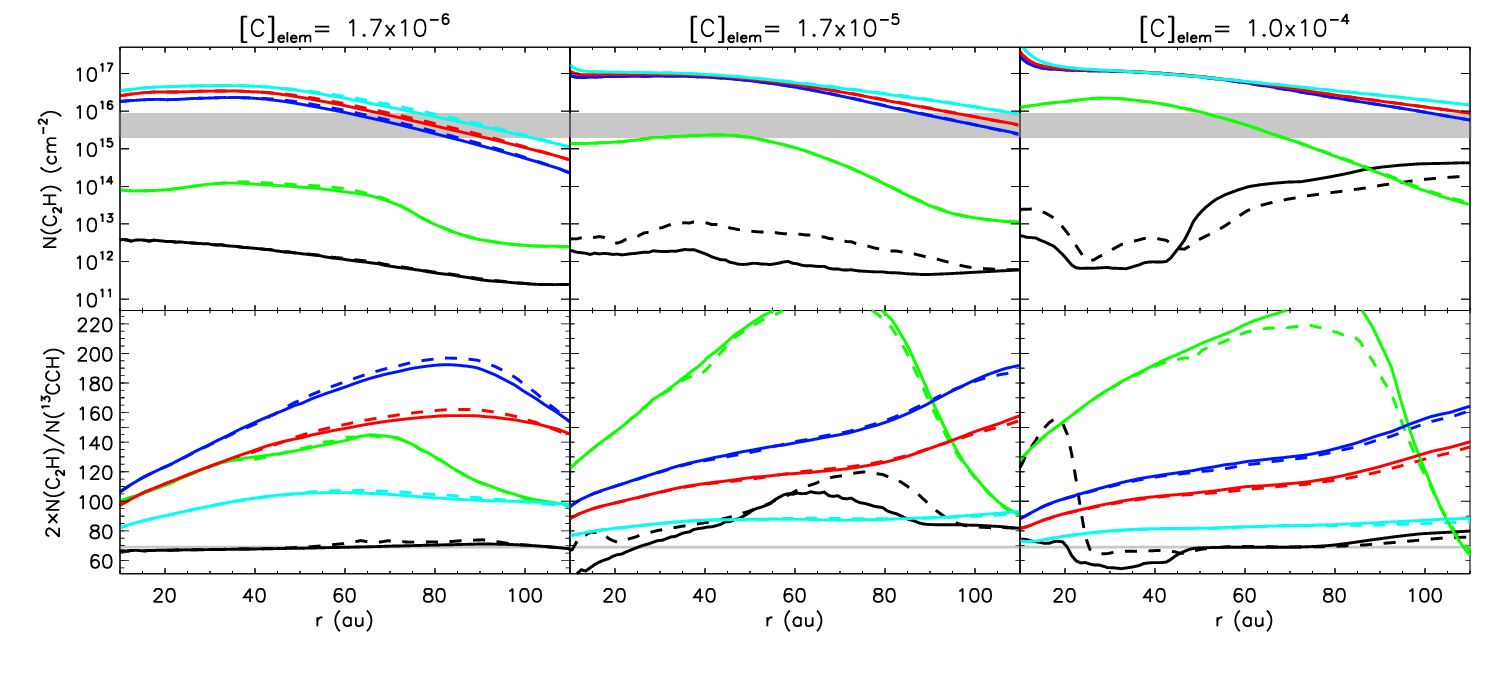}
\caption{Same as Figure~\ref{fig:all_co} except for C$_2$H. The gray boxes indicate the observed C$_2$H column density \citep{Kastner2014}. }
\label{fig:all_C$_2$H}
\end{figure*}

The bright C$_2$H emission in the disk indicates that the \ector\,ratio exceeds unity. The observed C$_2$H column density toward the TW Hya disk \citep{Kastner2014} could be fit with the models with \ector\,$\ge$\,1.5 as shown in the previous works \citep[e.g.,][]{Bergin2014,Lee2021}. However, the \cchqcr\,ratio toward TW Hya has been not reported yet. Thus, we expected the \cchqcr\,ratio based on our model. Note that $^{13}$CCH and C$^{13}$CH are treated as the same species in our work. Their carbon isotope ratios could be differ through $^{13}$CCH + H $\rightarrow$ C$^{13}$CH + H + 8.1 K, thus, their isotope ratios are similar in our work \citep{Furuya2011}.

Figure~\ref{fig:all_C$_2$H} shows the C$_2$H  column densities (top panels) and the column density ratios of C$_2$H and C$^{13}$CH (bottom panels).    When \cele\ is 1.7\,$\times$\,10$^{-6}$, the \cchqcr\,ratio increases up to $\sim$180 at the \ector\,ratio of 1.5 (blue line) then it decreases with increasing the \ector\,ratio. The \cchqcr\,ratio is governed by a balance between the inheritance from \cpqcr\ (resulting in an elevated \cchqcr) and the reactions (\ref{eq:6}) and (\ref{eq:7}) (leading to a reduction in the \cchqcr). Thus, the maximum \cchqcr\,ratio ($\sim$200) is achieved at the \ector\,ratio of $\sim$1 when \cele\ is 1.7\,$\times$\,10$^{-5}$ and 1.0\,$\times$\,10$^{-4}$ because of the growing significance of the latter contribution.

\subsection{Gas phase carbon isotope ratio}

\begin{figure*}[t]
\includegraphics[width=0.9\textwidth]{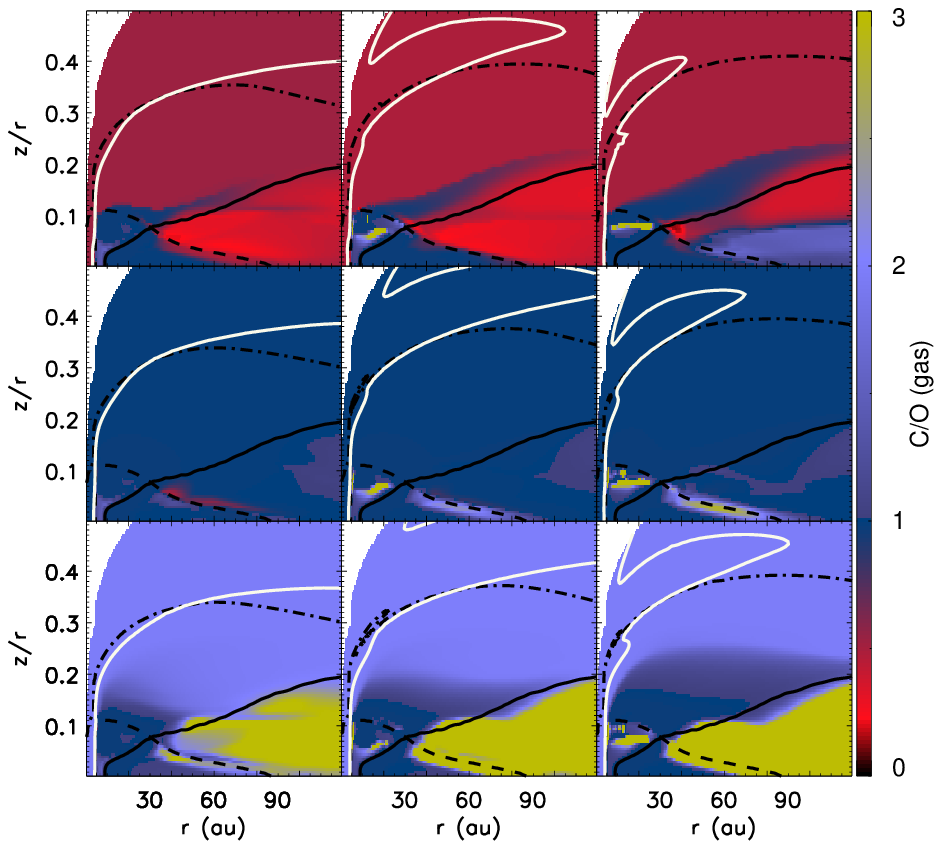}
\caption{Distribution of gas phase \ector\,ratio. Models have   \cele\,=\,1.7\,$\times$\,10$^{-6}$ (left), 1.7\,$\times$\,10$^{-5}$ (middle), and 1.0\,$\times$\,10$^{-4}$ (right) and the \ector\,ratios of 0.5, 1.0, and 2.0 from top to bottom layers. The solid and dashed black curves indicate the CO snow surface using the balancing between freeze-out onto dust grains and thermal evaporation (20--30\,K), and the dust-attenuated UV flux of 0.1 Draine Field \citep{Draine1978}. The dot-dashed lines represent the CO-photodissociation front where the CO abundance is a half of \cele. The solid white line indicate the gas temperature of 100\,K. }
\label{fig:tot_c2o}
\end{figure*}

The isotope ratio measured in disks is primarily in gas phase carriers. In this section we focus on its gas-phase values for different molecules commonly observed.
The \coqcr\,ratio depends on the gas-phase \ector\,ratio (hereafter \gctor) while the \ector\, ratio mentioned in this work is the total values including both gas and ice phases. Figure~\ref{fig:tot_c2o} shows the 2\,D distribution of only the \gctor\,ratios. The \gctor\,ratio in the warm molecular layer is close to the total value (\ector) except for the lower boundary where the \gctor\,ratio is close to unity. 
When the \ector\,ratio is 0.5 (top rows), CO$_2$ ice is dominant just above the CO snow surface, thus, the \gctor\,ratio approaches unity. When the \ector\,ratio is higher than unity (bottom rows), HCN and carbon chain ices, like C$_m$H$_n$, attain notable abundances (comparable to CO) near the lower boundary of the warm molecular layer in the inner disk. This occurrence leads to the \gctor\,ratio approaching unity regardless of the \ector\, ratio and \cele.
However, in mid-plane at the outer disk, gas phase CO freezes out onto dust grains below the CO snow surface, and atomic carbon is dominant in the gas phase. Therefore, the \gctor\,ratio is higher than the \ector\,ratio when the \ector\,ratio exceeds unity, while the \gctor\,ratio is close to the \ector\,ratio in other cases regardless of \cele. 

\begin{figure*}[t]
\includegraphics[width=0.9\textwidth]{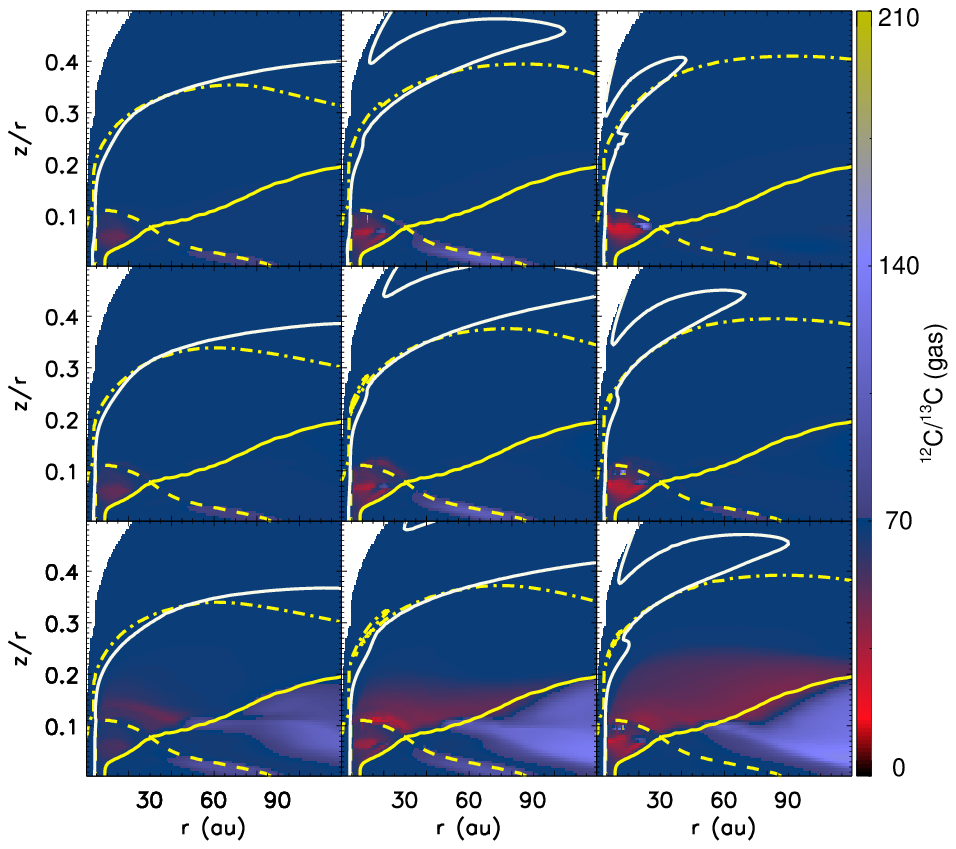}
\caption{Same as Figure~\ref{fig:tot_c2o} except for the gas phase $^{12}$C/$^{13}$C isotope ratio. The colors of the curves are changed for better visibility. }
\label{fig:tot_iso_gas}
\end{figure*}

Figure~\ref{fig:tot_iso_gas} shows the gas phase carbon isotope ratio. The gas phase carbon isotope ratio exhibits  a similar trend independent of  \cele. When the \ector\,ratio is $\leq$\,1, the carbon isotope ratio is close to \ecqcr\ above the UV flux of 0.1 Draine field (the dashed yellow lines). However, when the \ector\,ratio exceeds unity, $^{13}$C is enriched in the gas phase near the lower boundary of the warm molecular layer where CO is abundant in the gas phase (above the CO snow surface and the UV flux of 0.1 Draine field). It implies that ice abundances are comparable to the CO abundance and  $^{13}$C is deficient in the ice phase there. On the other hand, the gas exhibits a $^{13}$C-deficiency below the CO snow surface in the outer disk where only small portion ($\sim$a few\,\%) of elemental carbon is in the gas phase and dominant gas phase species are atomic carbon. 

\subsection{Effects of Ionization rate }\label{sec:eoi}

\begin{figure*}[t]
\includegraphics[width=1.0\textwidth]{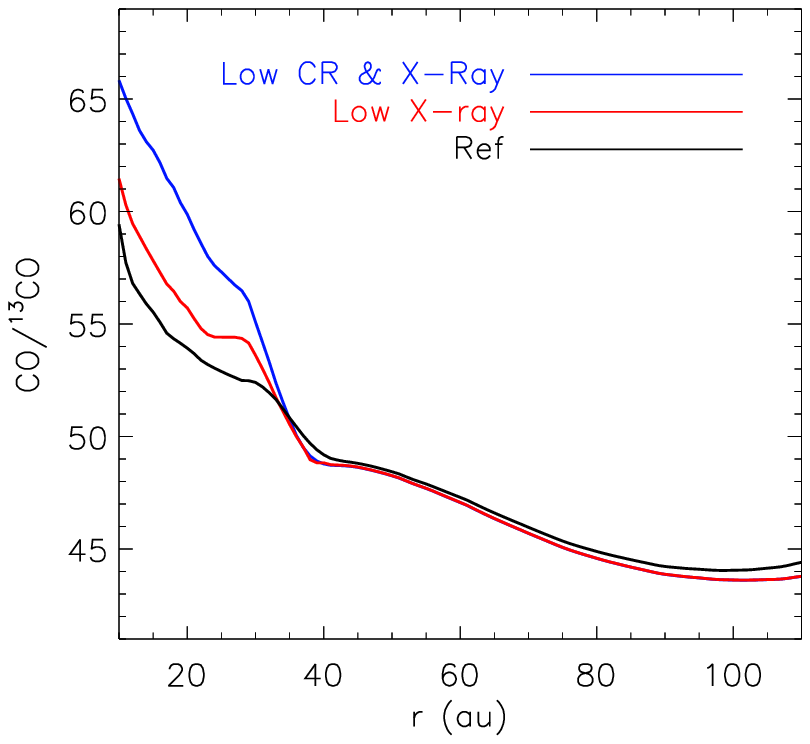}
\caption{Effects of ionization rates on the \coqcr\,ratio. The reference model is represented in the black line. The red and  blue lines indicate the models with a X-ray luminosity 10$^{-3}$ times that in the reference model, and additionally, a cosmic ray ionization rate one tenth of  the reference model's value, respectively.}
\label{fig:effect1}
\end{figure*}

Carbon fractionation mainly occurs in the cold regions through the isotope-exchange reactions \citep[e.g.,][]{Nomura2023}. 
Thus, ionization rates might affect the carbon isotope ratios. In the lower boundary of the warm molecular layer, the formation of  C$^+$ and $^{13}$C$^+$ starts from the reaction of CO\,+\,He$^+$ and the later is ionized by X-rays and/or cosmic rays in the reference model.  Therefore, the efficiency of the isotope exchange reaction (\ref{eq:1}) might depend on the ionization rates by X-rays and/or cosmic rays.  

The reference model could fit the observations of the ionization tracers, HCO$^+$ and N$_2$H$^+$. The HCO$^+$ column density, which is derived by using HC$^{18}$O$^+$ 4--3 line \citep{Furuya2022b}, is fitted with the reference model as mentioned in Section~\ref{subsec:hcop} although the HCO$^+$ column density is insensitive to the ionization rates \citep{Aikawa2021}. Our model could also fit the observed N$_2$H$^+$ column density of $\sim$10$^{13}$ cm$^{-2}$, which is measured using N$_2$H$^+$ 1--0 and 4-3 lines \citep{Cleeves2015,Schwarz2019}. Note that \citet{Cleeves2015} determined an ionization rate (X-ray + CR) below 10$^{-19}$\,s$^{-1}$ at the disk midplane, derived through the fitting of the observed HCO$^+$ and N$_2$H$^+$ data. In contrast, our model shows an ionization rate exceeding 10$^{-18}$\,s$^{-1}$, as presented in the bottom left panel of Figure~\ref{fig:phys}. 
We find a reduced ionization rate is needed, however how the \ector\,ratio and the isotope ratio of carbon may play an important role that should be explored in future work.

Figure~\ref{fig:effect1} shows the effects of ionization rates to \coqcr\,ratio. In the reference model (the black line), the ionization by X-rays is more dominant than that by cosmic rays \citep[with the rate of 5\,$\times$\,10$^{-19}$\,s$^{-1}$][]{Kama2016} except for the mid-plane of the inner disk ($<$\,10\,au). The red and blue lines indicate the model with X-ray luminosity 10$^{-3}$ times that of the reference model ($L_X$). The latter model has an order of magnitude lower cosmic ray ionization rate compared to the reference model. The ionization rates at $z/r$\,=\,0.1, where the contribution to the column density is maximum, at $r$\,=\,30\,au are 3.2\,$\times$\,$10^{-18}$\,s$^{-1}$, 4.7\,$\times$\,$10^{-19}$\,s$^{-1}$, and 5.0\,$\times$\,$10^{-20}$\,s$^{-1}$, for the black, red and blue line models, respectively.
The column densities of HCO$^+$ are similar within a factor of two among the models while the N$_2$H$^+$ column densities in the red and blue line models are a factor of $\sim$10 and $\sim$30 lower than that in the reference model.

\section{Summary}\label{sec:summary}
We have investigated the effects of elemental carbon to oxygen (\ector) ratio  and elemental carbon abundance to hydrogen (\cele) on the carbon isotope ratio in the disk. 

1. The \coqcr\, ratio monotonically decreases with increasing \ector\, ratio, and CO appears 25\% and $>$50\% more fractionated in the outer disk ($>$40\,au) when \ector\, are 1.5 and 5, respectively. The \coqcr ratio is close to \ecqcr
when \ector\, is $\le$\,1 except for the inner disk ($<$20\,au) regardless of \cele. 
When \ector\,$>$\,1, CO is enriched in $^{13}$C in the warm molecular layer through the isotope exchange reaction (\ref{eq:1}).  The \coqcr\,ratio observed in the TW Hya disk could be reproduced when the \ector\,ratio is higher than 2--5.

2. \hcopqcr\ ratio is by $\sim$20\% more fractionated compared to \coqcr\, ratio in the inner disk ($<$40\,au) regardless of \ector. Most models could reproduce the observed \hcopqcr\ ratio of 54$\pm$17 within the error range.
 
3. It is not straightforward to infer the \ector\,ratio from the carbon isotope ratios of HCN and carbon chain molecules  (e.g., C$_2$H). 
This is because they exhibit an increasing trend followed by a decrease according to the \ector\, ratio. In addition, they are also affected by \cele. 
However, their column densities are sensitive to \ector\, ratio, and thus, can be used as an indicator of \ector\, ratio. Thus, chemical models are essential for interpreting observations. 
The \hcnqcr\,ratio observed in the TW Hya disk could be reproduced when the \ector\,ratio of 2--5.

4. When \ector\,$>$\,1, in the vicinity of the lower boundary of warm molecular layer, the $^{13}$C-enriched CO dominates in the gas phase. In addition, the gas phase \ector\,ratio approaches unity because excess elemental carbon in the gas phase is removed through ice formation in the form of HCN and carbon chain molecules (C$_m$H$_n$), which exhibit $^{13}$C depletion. Thus, $^{13}$C is enriched in the bulk of the gas while being deficient in the bulk of the ice phase. 

\section*{Acknowledgements}
We are grateful to the anonymous referee for the valuable comments that helped to improve the manuscript. This work is supported by the Korea Astronomy and Space Science Institute under the R\&D program (Project No. 2024-1-841-00) supervised by the Ministry of Science and ICT. 
H.N. acknowledges financial support by JSPS and MEXT Grants-in-Aid for Scientific Research, 18H05441, 19K03910, 20H00182.
K.F. acknowledges financial support by JSPS and MEXT Grants-in-Aid for Scientific Research, 20H05847 and 21K13967.

\bibliography{main}{}
\bibliographystyle{aasjournal}

%% This command is needed to show the entire author+affiliation list when
%% the collaboration and author truncation commands are used.  It has to
%% go at the end of the manuscript.
%\allauthors

%% Include this line if you are using the \added, \replaced, \deleted
%% commands to see a summary list of all changes at the end of the article.
%\listofchanges

\clearpage

\appendix
\restartappendixnumbering 

\section{2 D distributions of abundance and carbon isotope ratio}

\begin{figure*}[h]
\includegraphics[width=0.9\textwidth]{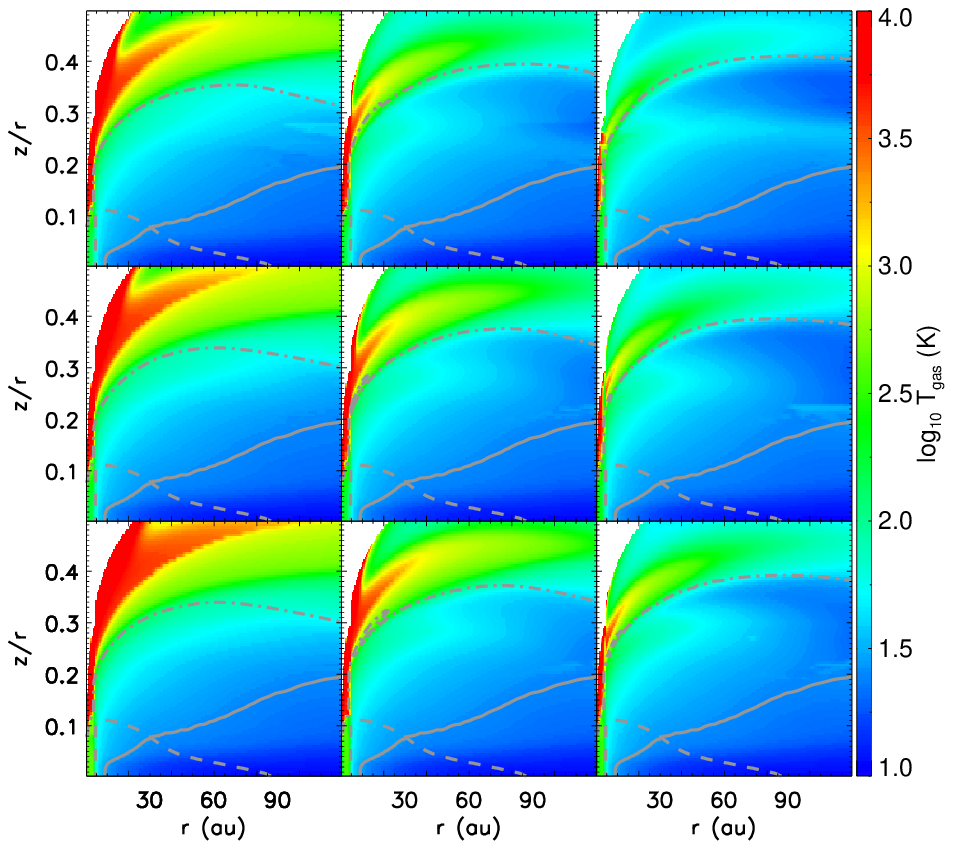}
\caption{Gas temperature distributions for the models with  \cele\,=\,1.7\,$\times$\,10$^{-6}$ (left), 1.7\,$\times$\,10$^{-5}$ (middle), and 1.0\,$\times$\,10$^{-4}$ (right).
The \ector\,ratios of 0.5, 1.0, and 2.0 are presented from top to bottom layers. The solid and dashed gray curves indicate the CO snow surfaces using the balancing between freeze-out onto dust grains and thermal evaporation (20--30\,K), and the dust-attenuated UV flux of 0.1 Draine Field \citep{Draine1978}. The dot-dashed lines represent the CO-photodissociation front where the CO abundance is a half of \cele.}
\label{fig:all_2d_tk}
\end{figure*}

\begin{figure*}[t]
\includegraphics[width=0.9\textwidth]{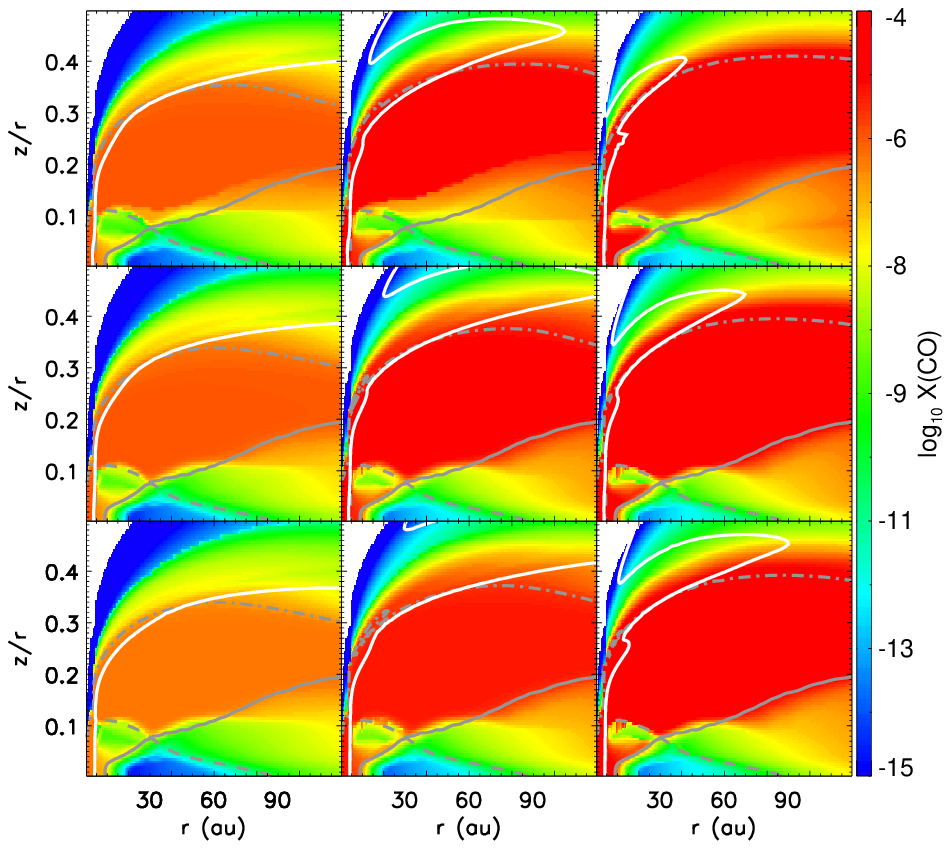}
\caption{CO abundance distributions for the same models in Figure~\ref{fig:all_2d_tk}. The gray curves are the same as those in Figure~\ref{fig:all_2d_tk}. The solid white curve indicates the gas temperature of 100~K.}
\label{fig:all_2d_co}
\end{figure*}

\begin{figure*}[t]
\includegraphics[width=0.9\textwidth]{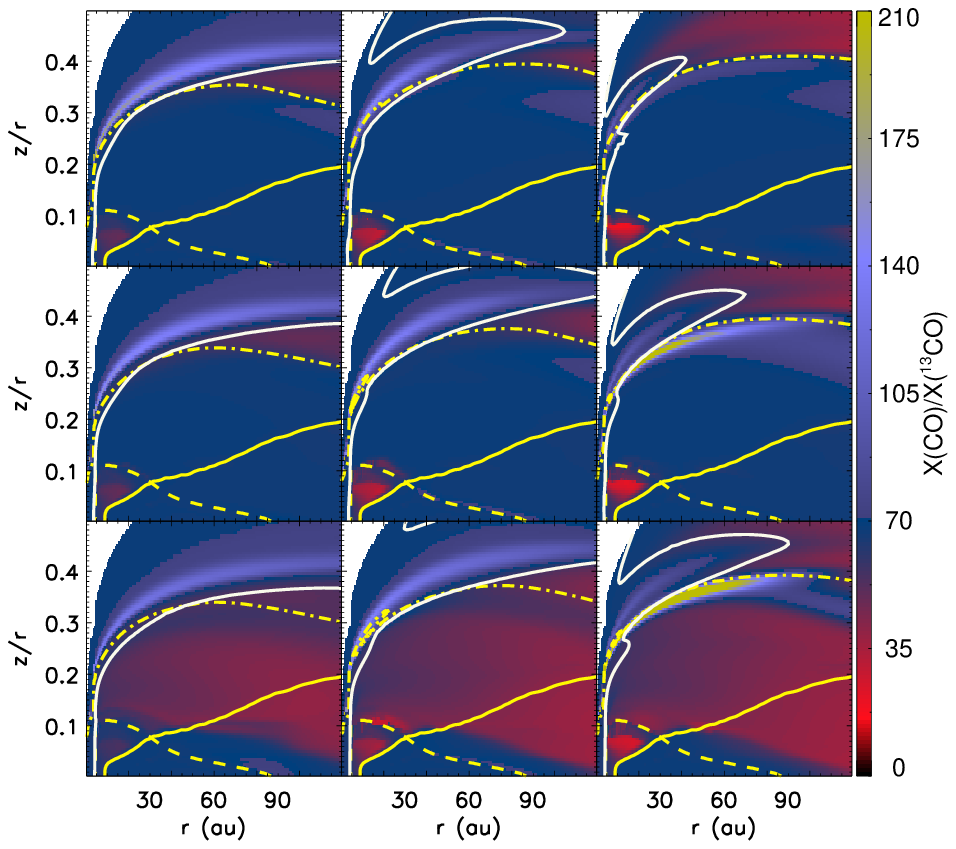}
\caption{CO/$^{13}$CO isotope ratio distributions for the same models in Figure~\ref{fig:all_2d_co}. }
\label{fig:all_2d_iso_co}
\end{figure*}

\begin{figure*}[t]
\includegraphics[width=0.9\textwidth]{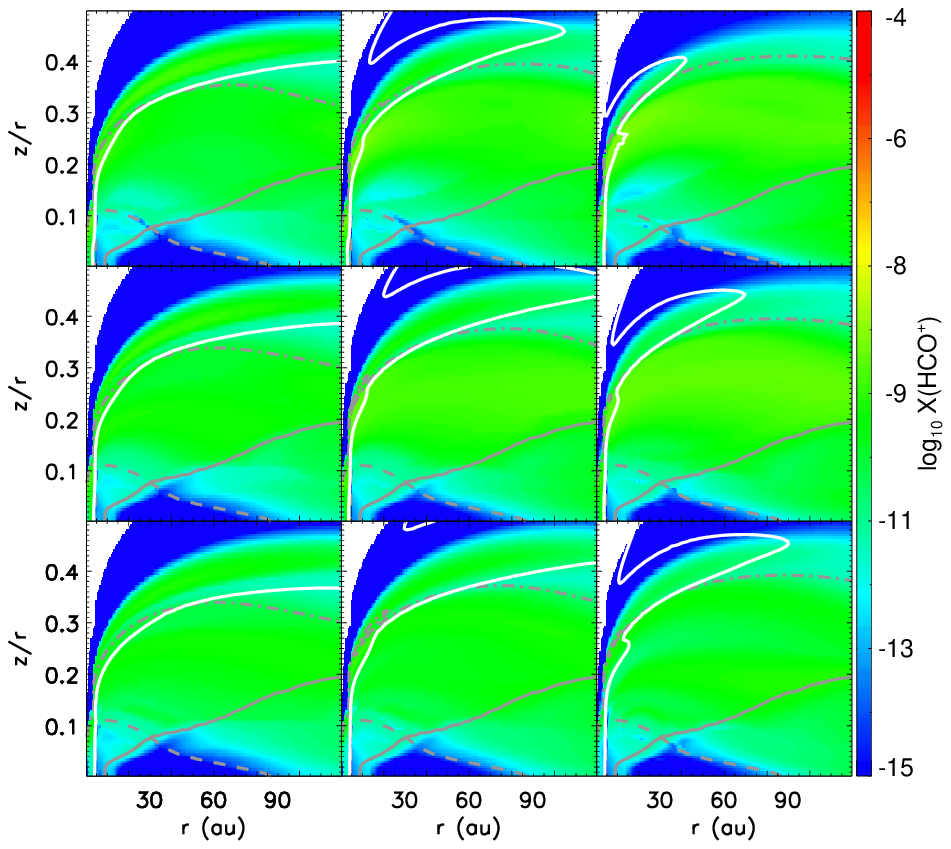}
\caption{Same as Figure~\ref{fig:all_2d_co} except for HCO$^+$.}
\label{fig:all_2d_hcop}
\end{figure*}

\begin{figure*}[t]
\includegraphics[width=0.9\textwidth]{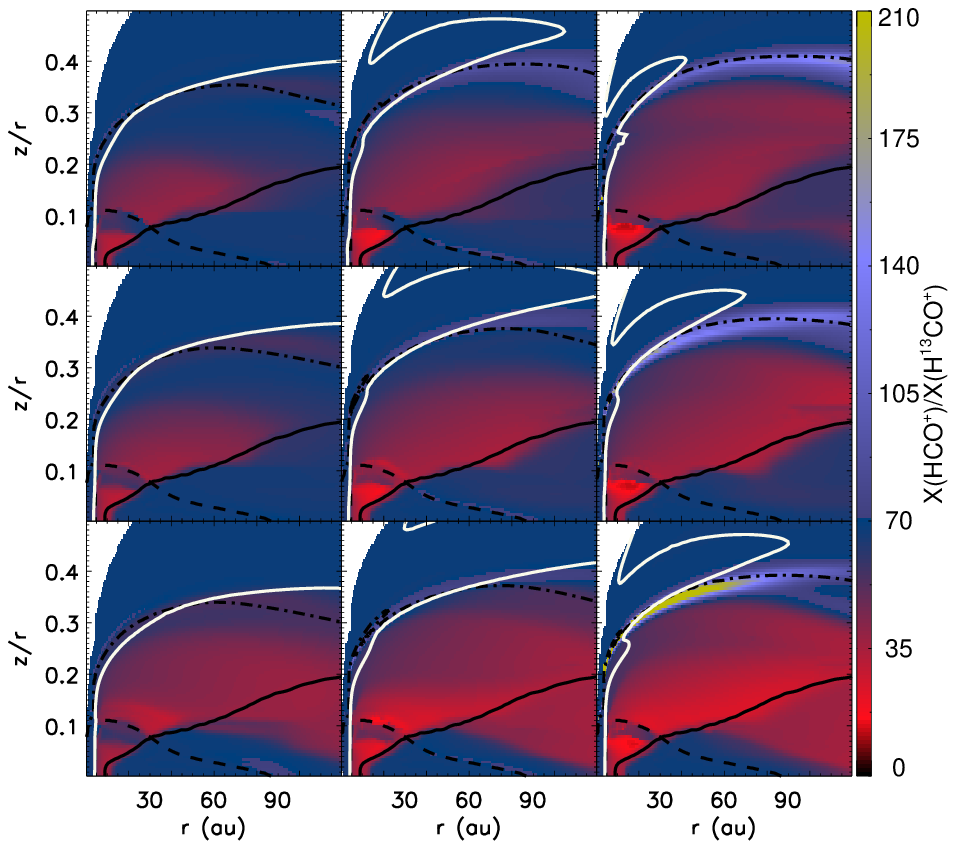}
\caption{Same as Figure~\ref{fig:all_2d_iso_co} except for HCO$^+$/H$^{13}$CO$^+$ isotope ratio.}
\label{fig:all_2d_iso_hcop}
\end{figure*}

\begin{figure*}[t]
\includegraphics[width=0.9\textwidth]{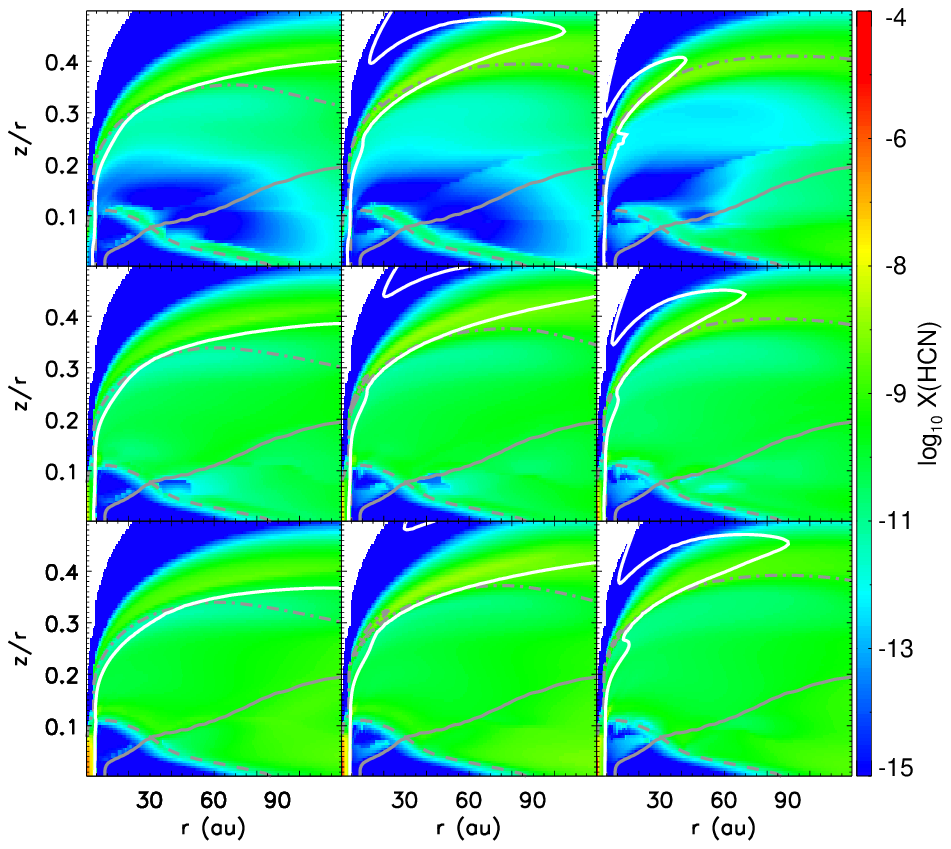}
\caption{Same as Figure~\ref{fig:all_2d_co} except for HCN.}
\label{fig:all_2d_hcn}
\end{figure*}

\begin{figure*}[t]
\includegraphics[width=0.9\textwidth]{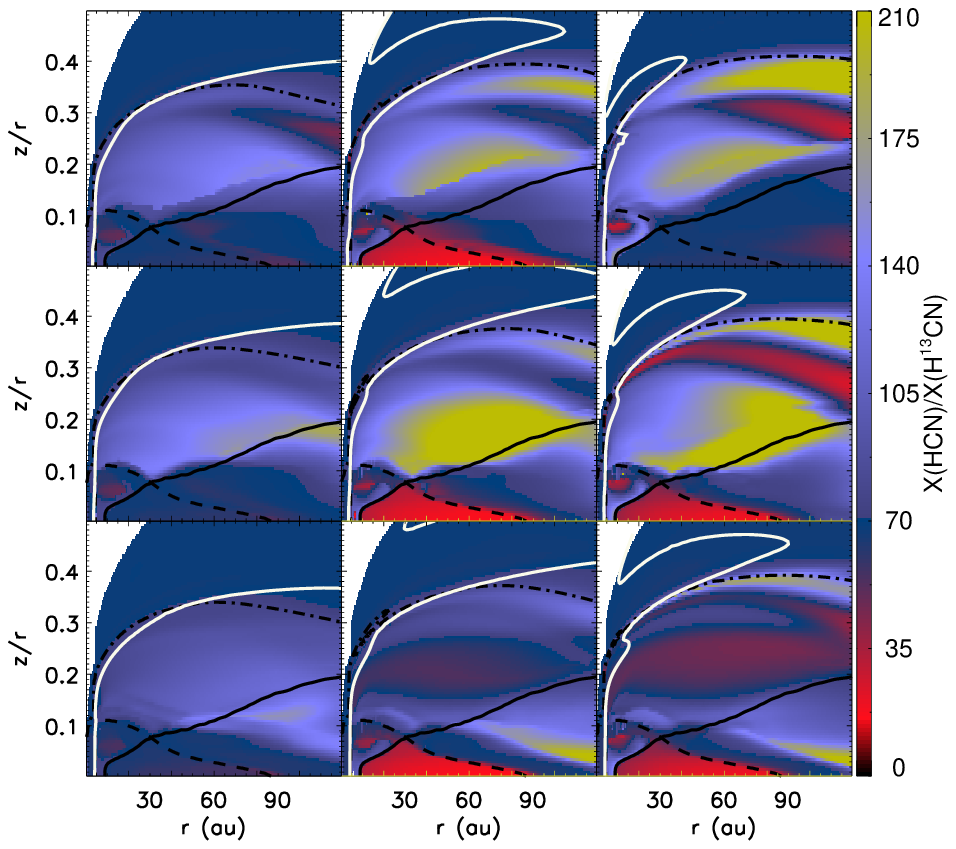}
\caption{Same as Figure~\ref{fig:all_2d_iso_co} except for HCN/H$^{13}$CN isotope ratio.}
\label{fig:all_2d_iso_hcn}
\end{figure*}

\begin{figure*}[t]
\includegraphics[width=0.9\textwidth]{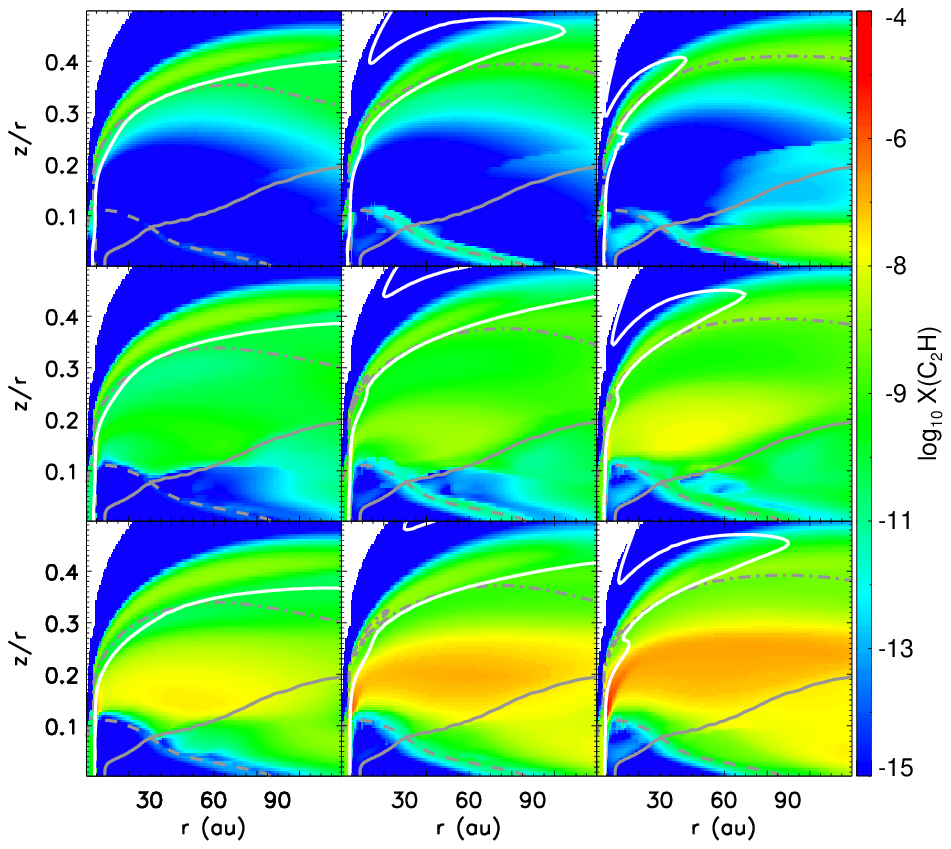}
\caption{Same as Figure~\ref{fig:all_2d_co} except for C$_2$H.}
\label{fig:all_2d_C$_2$H}
\end{figure*}

\begin{figure*}[t]
\includegraphics[width=0.9\textwidth]{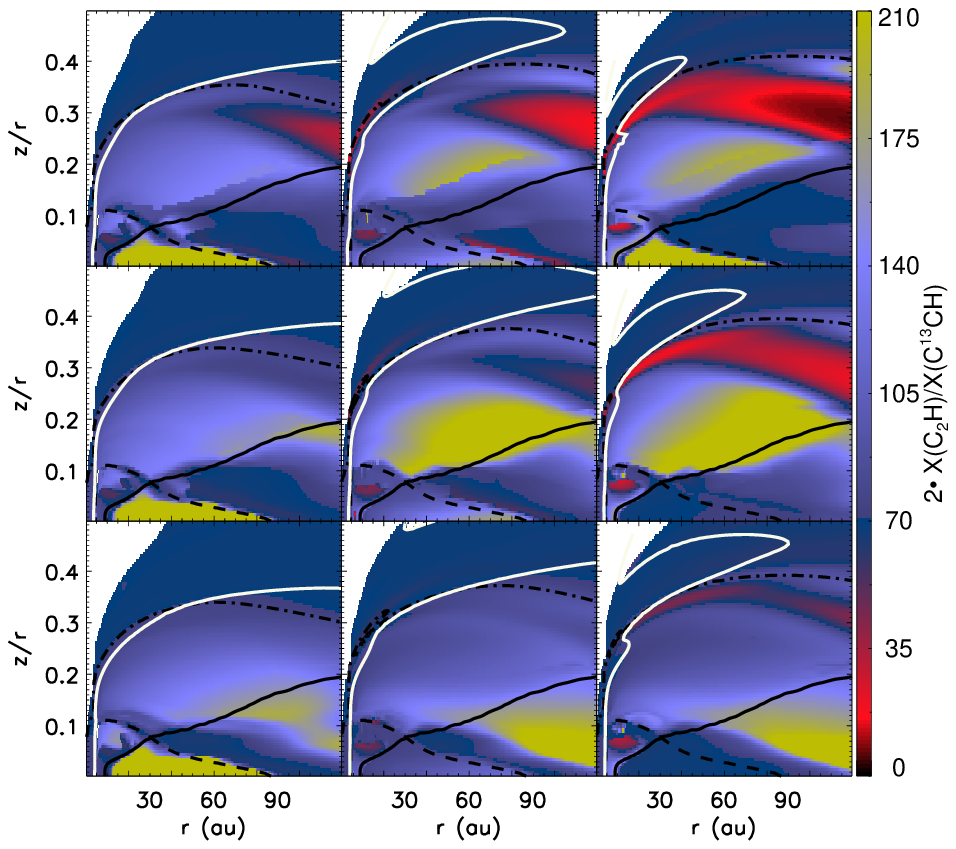}
\caption{Same as Figure~\ref{fig:all_2d_iso_co} except for C$_2$H/C$^{13}$CH isotope ratio.}
\label{fig:all_2d_iso_C$_2$H}
\end{figure*}
\clearpage
\end{document}